\documentclass{IEEEtran}

\usepackage[switch]{lineno}
\usepackage{amsmath,amssymb,amsfonts}
\usepackage{graphicx}
\usepackage{multirow}
\usepackage{textcomp,nicefrac}
\usepackage{bm}
\usepackage[dvipsnames]{xcolor}
\usepackage{hyperref}
\hypersetup{
    colorlinks=true,
    linkcolor=red,
    urlcolor=blue,
    citecolor=ForestGreen,
}
\usepackage{url}
\usepackage{cite}

\usepackage{ifthen} 
\newboolean{uprightparticles}
\setboolean{uprightparticles}{false} 
\usepackage{xspace} 
\usepackage{upgreek}

\def\MagUp {\mbox{\em Mag\kern -0.05em Up}\xspace}

\ifthenelse{\boolean{uprightparticles}}%
{

 \def\PDelta      {\ensuremath{\Delta}\xspace}                 
 \def\PXi         {\ensuremath{\Xi}\xspace}                 
 \def\PLambda     {\ensuremath{\Lambda}\xspace}                 
 \def\PSigma      {\ensuremath{\Sigma}\xspace}                 
 \def\POmega      {\ensuremath{\Omega}\xspace}                 
 \def\PUpsilon    {\ensuremath{\Upsilon}\xspace}
 \let\oldPi\Pi
 \def\PPi         {\ensuremath{\oldPi}\xspace}

 \def\PB      {\ensuremath{\mathrm{B}}\xspace}                 
                  
 \def\PD      {\ensuremath{\mathrm{D}}\xspace}

 \def\PK      {\ensuremath{\mathrm{K}}\xspace}

 \def\Pi      {\ensuremath{\mathrm{i}}\xspace}

 \def\Ps      {\ensuremath{\mathrm{s}}\xspace}

 \def\thebaroffset{0.0em}
}
{

 \mathchardef\PDelta="7101
 \mathchardef\PXi="7104
 \mathchardef\PLambda="7103
 \mathchardef\PSigma="7106
 \mathchardef\POmega="710A
 \mathchardef\PUpsilon="7107
 \mathchardef\PPi="7105
                  
 \def\PB      {\ensuremath{B}\xspace}                 
                  
 \def\PD      {\ensuremath{D}\xspace}

 \def\PK      {\ensuremath{K}\xspace}

 \def\Pi      {\ensuremath{i}\xspace}

 \def\Ps      {\ensuremath{s}\xspace}

 \def\thebaroffset{0.18em}
}
\newcommand{\offsetoverline}[2][\thebaroffset]{\kern #1\overline{\kern -#1 #2}}%

\makeatletter
\ifcase \@ptsize \relax
  \newcommand{\miniscule}{\@setfontsize\miniscule{4}{5}}
\or
  \newcommand{\miniscule}{\@setfontsize\miniscule{5}{6}}
\or
  \newcommand{\miniscule}{\@setfontsize\miniscule{5}{6}}
\fi
\makeatother

\DeclareRobustCommand{\optbar}[1]{\shortstack{{\miniscule (\rule[.5ex]{1.25em}{.18mm})}
  \\ [-.7ex] $#1$}}

\def\squark    {{\ensuremath{\Ps}}\xspace}

\def\KorKbar {\kern \thebaroffset\optbar{\kern -\thebaroffset \PK}{}\xspace}

\def\D       {{\ensuremath{\PD}}\xspace}

\def\DorDbar {\kern \thebaroffset\optbar{\kern -\thebaroffset \PD}\xspace}

\def\Dp      {{\ensuremath{\D^+}}\xspace}
\def\Dm      {{\ensuremath{\D^-}}\xspace}

\def\DpDm    {\ensuremath{\Dp {\kern -0.16em \Dm}}\xspace}

\def\B       {{\ensuremath{\PB}}\xspace}

\def\BorBbar {\kern \thebaroffset\optbar{\kern -\thebaroffset \PB}\xspace}

\def\Bd      {{\ensuremath{\B^0}}\xspace}

\def\BdorBdbar {\kern \thebaroffset\optbar{\kern -\thebaroffset \Bd}\xspace}

\def\Bs      {{\ensuremath{\B^0_\squark}}\xspace}

\def\BsorBsbar {\kern \thebaroffset\optbar{\kern -\thebaroffset \Bs}\xspace}

\def\Y#1S{\ensuremath{\PUpsilon{(#1S)}}\xspace}

\def\LorLbar     {\kern \thebaroffset\optbar{\kern -\thebaroffset \PLambda}\xspace}

\def\to                 {\ensuremath{\rightarrow}\xspace}

\def\AT#1     {\ensuremath{A_{\mathrm{T}}^{#1}}\xspace}           

\def\C#1      {\ensuremath{\mathcal{C}_{#1}}\xspace}                       
\def\Cp#1     {\ensuremath{\mathcal{C}_{#1}^{'}}\xspace}                    
\def\Ceff#1   {\ensuremath{\mathcal{C}_{#1}^{\mathrm{(eff)}}}\xspace}        
\def\Cpeff#1  {\ensuremath{\mathcal{C}_{#1}^{'\mathrm{(eff)}}}\xspace}       
\def\Ope#1    {\ensuremath{\mathcal{O}_{#1}}\xspace}                       
\def\Opep#1   {\ensuremath{\mathcal{O}_{#1}^{'}}\xspace}                    

\newcommand{\aunit}[1]{\ensuremath{\text{\,#1}}}       

\newcommand{\tev}{\aunit{Te\kern -0.1em V}\xspace}
\newcommand{\gev}{\aunit{Ge\kern -0.1em V}\xspace}
\newcommand{\mev}{\aunit{Me\kern -0.1em V}\xspace}
\newcommand{\kev}{\aunit{ke\kern -0.1em V}\xspace}
\newcommand{\ev}{\aunit{e\kern -0.1em V}\xspace}
 
\newcommand{\mevc}{\ensuremath{\aunit{Me\kern -0.1em V\!/}c}\xspace}
\newcommand{\gevc}{\ensuremath{\aunit{Ge\kern -0.1em V\!/}c}\xspace}
\newcommand{\mevcc}{\ensuremath{\aunit{Me\kern -0.1em V\!/}c^2}\xspace}
\newcommand{\gevcc}{\ensuremath{\aunit{Ge\kern -0.1em V\!/}c^2}\xspace}

\def\mhz  {\ensuremath{\aunit{MHz}}\xspace}
\def\khz  {\ensuremath{\aunit{kHz}}\xspace}

\def\gsim{{~\raise.15em\hbox{$>$}\kern-.85em
          \lower.35em\hbox{$\sim$}~}\xspace}
\def\lsim{{~\raise.15em\hbox{$<$}\kern-.85em
          \lower.35em\hbox{$\sim$}~}\xspace}

\def\tell1  {TELL1\xspace}
\def\ukl1   {UKL1\xspace}

\def\BibTeX{{\rm B\kern-.05em{\sc i\kern-.025em b}\kern-.08em T\kern-.1667em\lower.7ex\hbox{E}\kern-.125emX}}
\begin{document}

\title{A FPGA-based architecture for real-time cluster finding in the LHCb silicon pixel detector}

\author{G. Bassi, L. Giambastiani, K. Hennessy, F. Lazzari, M. J. Morello, T. Pajero,  A. Fernandez Prieto,  G. Punzi

\thanks{Submitted on xx/yy/2023.}

\thanks{G. Bassi and M. J. Morello are with INFN Sezione di Pisa, Pisa, IT and Scuola Normale Superiore, Pisa, Italy (e-mail: giovanni.bassi@cern.ch).}

\thanks{L. Giambastiani was with INFN Sezione di Pisa, Pisa, IT and Università di Pisa, Pisa, IT. Now with Università degli Studi di Padova, Padova, Italy.}

\thanks{K. Hennessy is with the Department of Physics, Liverpool University, Liverpool L69 7ZE,
United Kingdom.}

\thanks{F. Lazzari is with INFN Sezione di Pisa, Pisa, IT and Università di Pisa, Pisa, Italy.}

\thanks{T. Pajero was with INFN Sezione di Pisa, Pisa, IT and Scuola Normale Superiore, Pisa, IT. Now with University of Oxford, Oxford OX1 3RH, United Kingdom.}

\thanks{A. Fernandez Prieto is with the Instituto Galego de Física de Altas Enerxías (IGFAE), Universidade de Santiago de Compostela, E-15782 Santiago de Compostela, Spain.}

\thanks{G. Punzi is with INFN Sezione di Pisa, Pisa, IT and Università di Pisa, Pisa, Italy.}}

\maketitle

\begin{abstract}
This article describes a custom VHDL firmware implementation of a two-dimensional cluster-finder architecture for reconstructing hit positions in the new vertex pixel detector (VELO) that is part of the LHCb Upgrade. This firmware has been deployed to the existing FPGA cards that perform the readout of the VELO, as a further enhancement of the DAQ system, and will run in real time during physics data taking, reconstructing VELO hits coordinates on-the-fly at the LHC collision rate. This pre-processing allows the first level of the software trigger to accept an 11\% higher rate of events, as the ready-made hit coordinates accelerate the track reconstruction and consumes significantly less electrical power. It additionally allows the raw pixel data to be dropped at the readout level, thus saving approximately 14\% of the DAQ bandwidth.
Detailed simulation studies have shown that the use of this real-time cluster finding does not introduce any appreciable degradation in the tracking performance in comparison to a full-fledged software implementation. This work is part of a wider effort aimed at boosting the real-time processing capability of HEP experiments by delegating intensive tasks to dedicated computing accelerators deployed at the earliest stages of the data acquisition chain.
\end{abstract}


\section{Introduction}
\label{sec:introduction}
\IEEEPARstart{T}{he} LHCb experiment has collected data over the past decade, during the Run\,1 and Run\,2 of the LHC, and recently underwent a major update for the current Run\,3.
In addition to replacing most of the subdetectors, the front-end electronics and data-acquisition system were completely renewed~\cite{online-upgrade}, to read out and process the complete information of the detector at the full LHC beam crossing rate of 40~MHz (30~MHz averaged over the LHC cycle). This change is motivated by the needs of the LHCb physics programme, which requires the collection of low transverse momentum events that need high-level processing to be distinguished from background events~\cite{upgrade2-eoi}.
This evolution puts a large computing toll on the new real-time processing system, motivating the deployment of innovative features, with a general trend of increasing customisation, parallelisation, and early data pre-processing. 
A new trigger system~\cite{online-upgrade, GPU-HLT} was designed to allow the experiment to collect data effectively at an instantaneous luminosity of $2 \times 10^{33}$~cm$^{-2}$s$^{-1}$, five times higher than during Run~2, corresponding to a bandwidth of about 32~Tb/s. The subsequent event-building stage and software high-level-trigger (HLT) processing lead to a data storage flow of 80~Gb/s. 

The triggering process is divided into two main stages, named HLT1 and HLT2. The HLT1 uses an array of GPU servers to perform a faster event reconstruction, with the only purpose of reducing the event rate, while retaining as much signal as possible, to a level acceptable for HLT2. The HLT2, based on an array of CPU servers, performs a complete reconstruction of events with an offline-level quality,  that is permanently stored for subsequent analysis. To perform its function effectively, the HLT1 needs to perform a nearly complete event reconstruction. First, it finds track segments in the VErtex LOcator detector (VELO), attaching to them hits from the further tracking stations upstream and downstream of the magnet to obtain complete tracks; then, the positions of the primary vertices of the proton-proton ($pp$) collisions are found, as well as those of displaced vertices that constitute the main signature of heavy-flavour particle decays.

The feasibility of implementing several parts of this sequence in a specialised architecture, using programmable digital electronics co-processors~(FPGAs), has been studied with the aim of achieving a faster and cheaper reconstruction, especially in view of future runs, moving parts of it before the event-building stage~\cite{cenci-twepp}. 

In this article, we address the very first step in the HLT1 event reconstruction, that is the search for clusters of active pixels in the VELO~\cite{velo-upgrade}.  Grouping contiguous pixels in clusters is a conceptually simple but computationally demanding task, due to the two-dimensional (2D) geometry and the large number of pixels of the VELO detector (approximately 40 million). In the preliminary version of the HLT1, designed to run entirely on CPUs, this task alone consumed 17\% of the time required by the complete HLT1 reconstruction sequence. We address here this issue and describe an efficient architecture of this functionality, requiring a very modest amount of FPGA resources, while providing the throughput and the performance required for its use within the LHCb DAQ system.
The core ideas underlying the design of this architecture are based on studies of an FPGA-based track-finding system, performed within the INFN-RETINA R\&D project~\cite{cenci-twepp}. 
The overall structure of our algorithm and its main building blocks are rather general, and can be applied to any pixel detector. 
A baseline version is available for download from a public code repository \cite{firmware-repo}. However, the LHCb version contains specific features tailored to the VELO detector.

\section{Format and features of VELO data}
\label{sec:format_velo}

\begin{figure}[tb]
    \centering
    \includegraphics[width=20 pc]{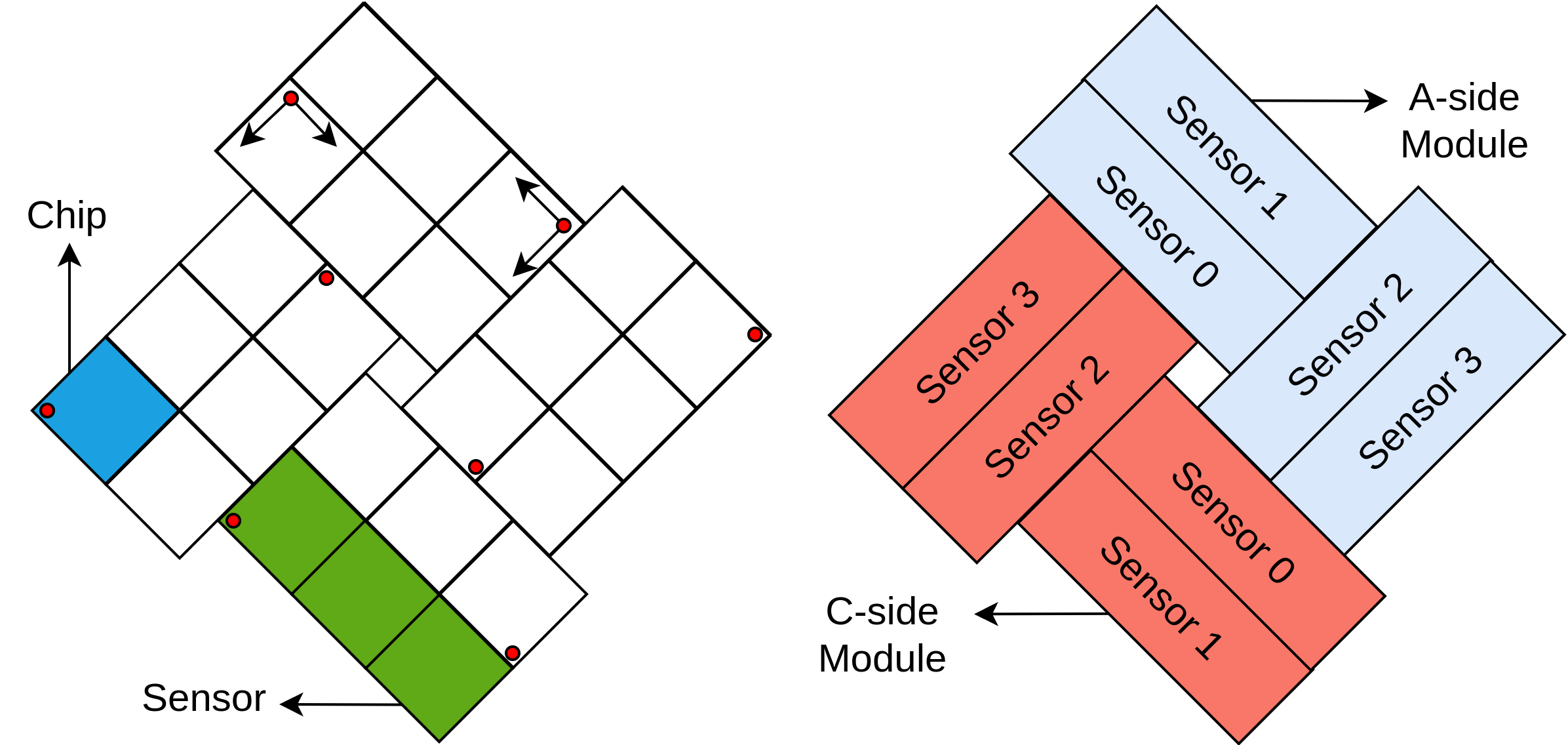}
    \caption{Illustration of the basic constituents of a VELO layer \cite{velo-upgrade}. Red dots mark the origin, pixel (0,0), of the local cartesian coordinate system of each sensor (see Sect.~\ref{sec:core_architecture}). As an example, the axis orientation is displayed in both sensors of the upper module.}
    \label{fig:layer-definitions}
\end{figure}

\begin{figure}[tb]
    \centering
    \includegraphics[width=20pc]{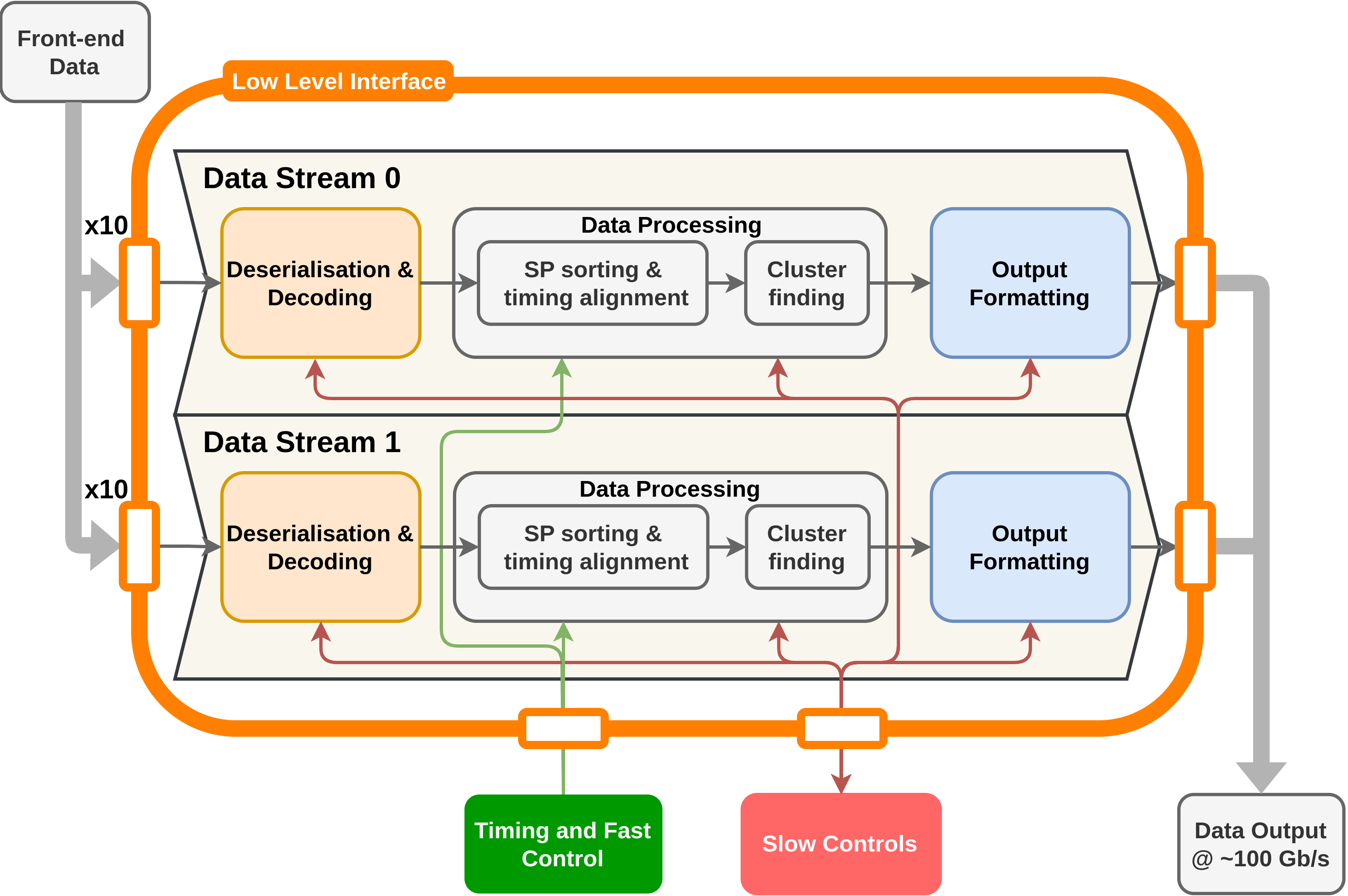}
    \caption{Schematics of the TELL40 firmware. A detailed description can be found in Ref.~\cite{velo-firmware}.}
    \label{fig:velo-firmware}
\end{figure}

The Run 3 VELO is a silicon pixel detector consisting of 26 layers both downstream (19) and upstream (7) of the nominal point of $pp$ collisions. Each layer consists of two modules, each read by a dedicated readout card. A module is made of four sensors, each of which is bump-bonded to three VeloPix \cite{velopix} ASICs (chips), as shown in~Fig.~\ref{fig:layer-definitions}.
The VELO front-end data arrive at the LHCb readout cards via optical links as aggregated groups of 4$\times$2 pixels, named SuperPixels (SPs), with binary response. 
Data are deserialized, decoded and sent to the data processing stage. SuperPixels are output by the detector without a well-defined time ordering, and data from different LHC beam crossings (separated by 25~ns) may not be synchronised and mixed over time. The first step of the VELO data-processing firmware reorders the SPs, making sure that SPs coming from the same proton bunch crossing (event) are grouped together~\cite{velo-firmware}, before data are sent to the clustering stage. Reconstructed clusters are then formatted into LHCb event fragments and sent to the PCIe bus.
Figure~\ref{fig:velo-firmware} shows a schematic view of the firmware~\cite{velo-firmware} of the custom PCIe cards (TELL40\cite{tell40}) that perform the readout of the VELO. TELL40 cards are used as readout units for each subdetector within LHCb. Each TELL40 card carries an Altera Arria-10 GX 1150 FPGA with 1150k logic elements.

The clustering firmware was designed to take as its input the list of all active SPs found in a given event, and to produce a list of reconstructed clusters, each with the local $(x,y)$ coordinates of its centroid. In addition, it provides the detailed shape of the pixel cluster, as well as some flags indicating cluster quality. These additional quantities are not required by HLT1 reconstruction, but are computed to allow the HLT2 to perform a fully optimised reconstruction of tracks, despite of the lack of the original raw pixel data.

\begin{figure}[tb]
    \centering
    \includegraphics[width=20pc]{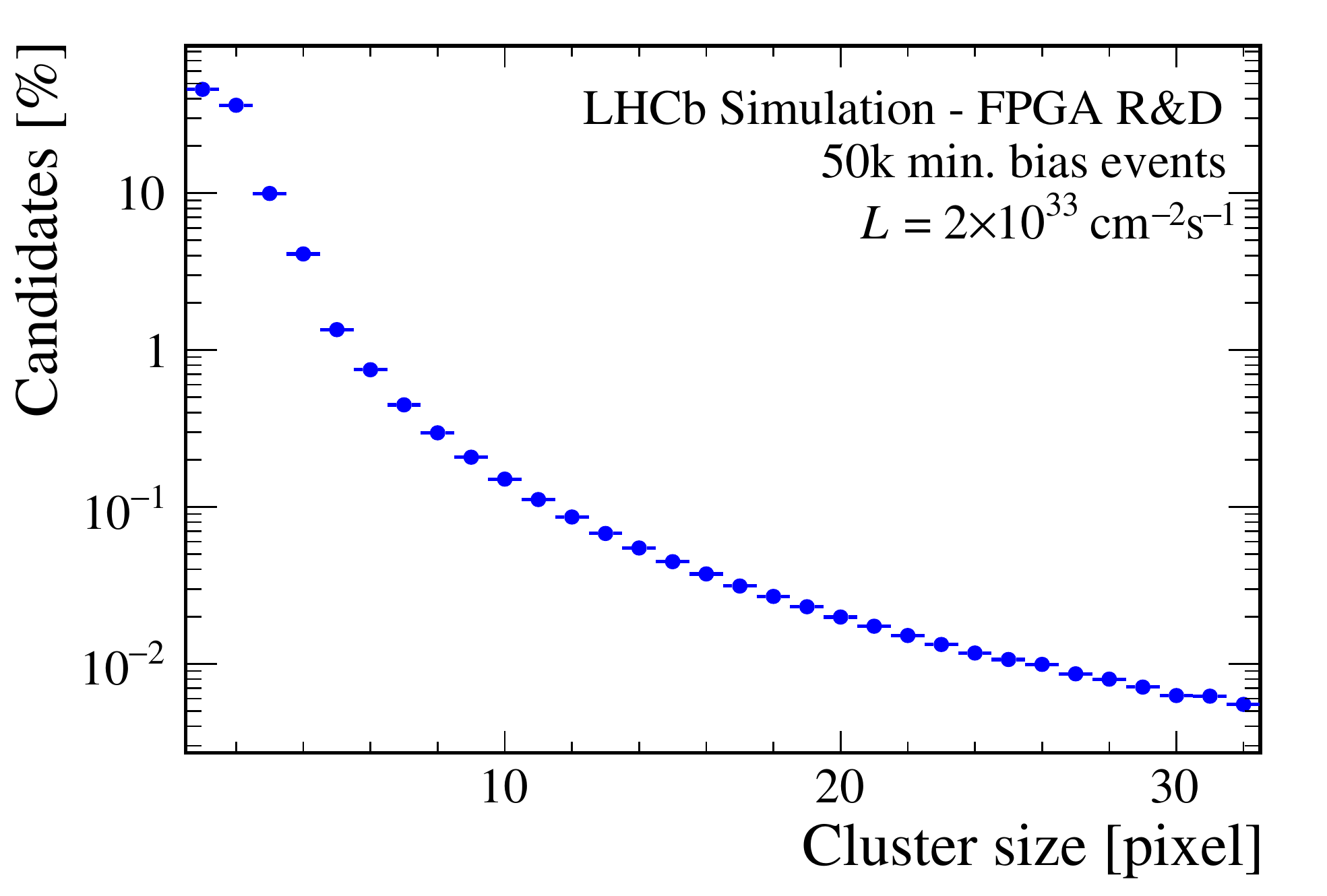}
    \caption{Distribution of the number of pixels per cluster.}
    \label{fig:cluster-size-ditribution}
\end{figure}

The size of clusters generated by individual charged particles crossing the VELO layers is less than or equal to 4 pixels in 96\% of the cases, whereas larger clusters are mostly the product of merged hits or secondary emissions ($\delta$-rays, etc.). The distribution of cluster sizes as predicted by the LHCb~Upgrade simulation (MC)~\cite{simulation} is shown in Fig.~\ref{fig:cluster-size-ditribution}.

\section{Core Architecture}
\label{sec:core_architecture}

The distribution in Fig.~\ref{fig:cluster-size-ditribution} implies that clusters produced by a single particle hit are often contained within a single SP word. In those cases, the reconstruction of the cluster can be performed through a look-up table, and it is therefore advantageous to perform an initial pre-processing of SPs to separate these occurrences  from the others, and to send them to  two distinct  parallel processing blocks. 
The separation is performed by comparing the 2D position of each SP with that of all other SPs of the same sensor in the same event. Each SP is then flagged as ``isolated'' if none of its eight SP neighbours has any active pixels.
The LHCb simulation predicts that isolated SPs will account for 53\% of the total number of SPs at nominal Run~3 luminosity conditions.

The centroid of clusters within isolated SPs is calculated directly by means of a look-up table (LUT). Each of the $2^8$ possible pixel configurations within a SP is linked to the precalculated centre of mass of the corresponding reconstructed cluster(s). 
This LUT-based reconstruction allows an extremely fast processing of isolated SPs, with a very limited amount of logic resources. 
It is possible for up to two distinct clusters to be present within a single SP. The firmware correctly handles this case as well, generating two independent clusters in the output.

\begin{figure}[tb]
    \centering
    \includegraphics[width=1\linewidth]{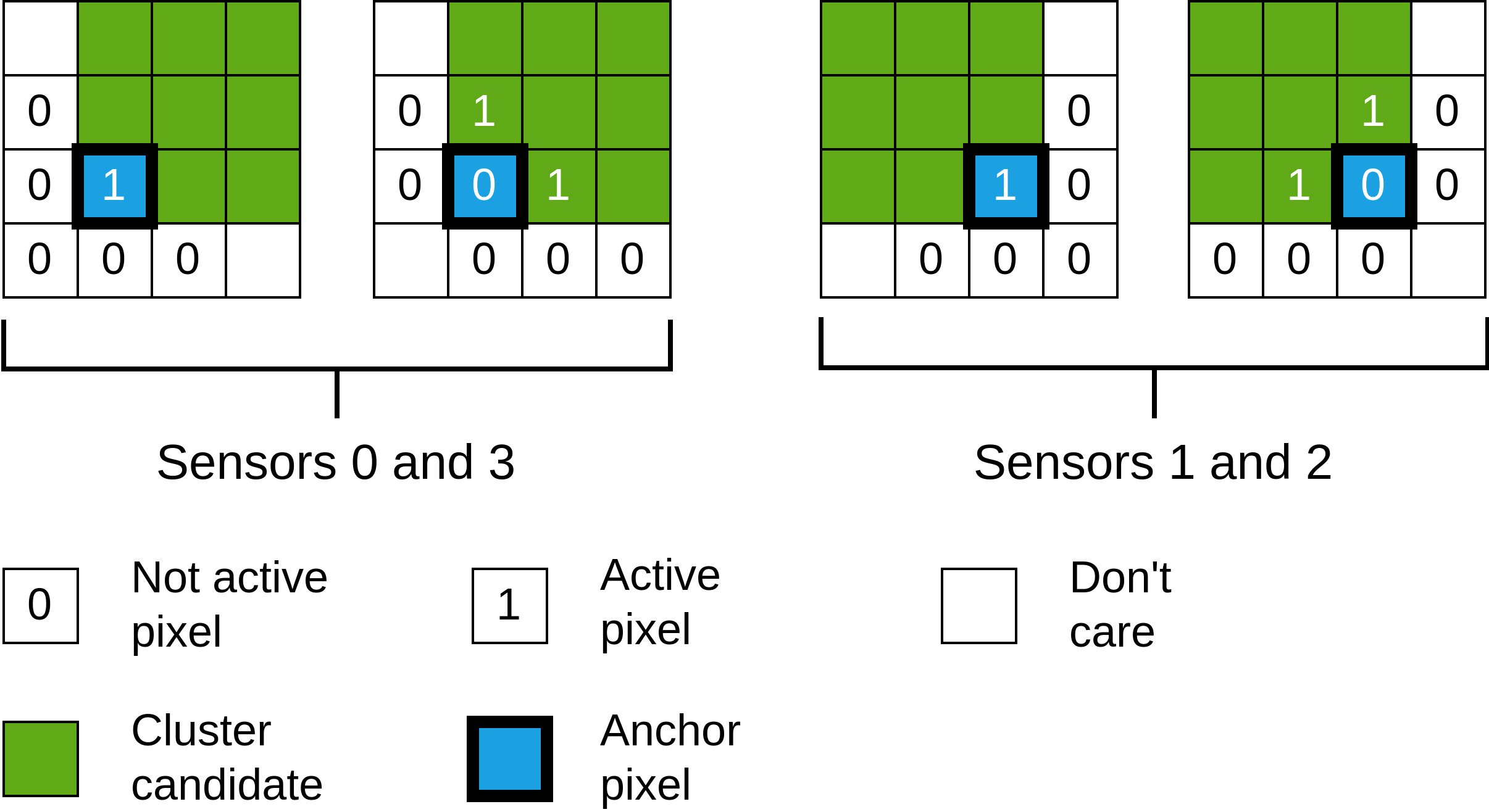}
    \caption{Pixel patterns used to identify a cluster candidate. The patterns are optimised for the sensor mounting orientation. See Ref.~\cite{luca-thesis, giovanni-thesis} for further details.}
    \label{fig:pixel-patterns}
\end{figure}

The algorithm for non-isolated SPs requires instead the concurrent processing of multiple SPs. This part of the processing is performed at the level of individual pixels, dropping the SP-based formatting of the data. Each detector pixel is mapped to a cell within an active bit matrix, set to 1 or 0 according to the pixel status. The bit matrix has a built-in logic, capable of recognising certain predetermined patterns, signalling the presence of a cluster corner at a certain pixel position. 
Since more than 96\% of the reconstructed clusters is made of no more than four contiguous pixels, the most efficient choice for the patterns is an ``L''~shaped sequence of inactive pixels with two different configurations of active pixels with the cluster candidate contained in a 3$\times$3 pixel grid (Fig.~\ref{fig:pixel-patterns}).
If one of the patterns is matched, the cluster candidate is recognised in the grid (green pixels in figure), as well as the anchor pixel (blue pixel in figure), positioned in one of the corners of the  grid depending on the orientation of the sensor.
The presence of a cluster corner is simultaneously checked on every bit of the matrix
and it is completed in a single clock cycle. In the next clock cycle, the first cluster found is extracted from the matrix.

\begin{figure}[tb]
    \centering
    \includegraphics[width=1\linewidth]{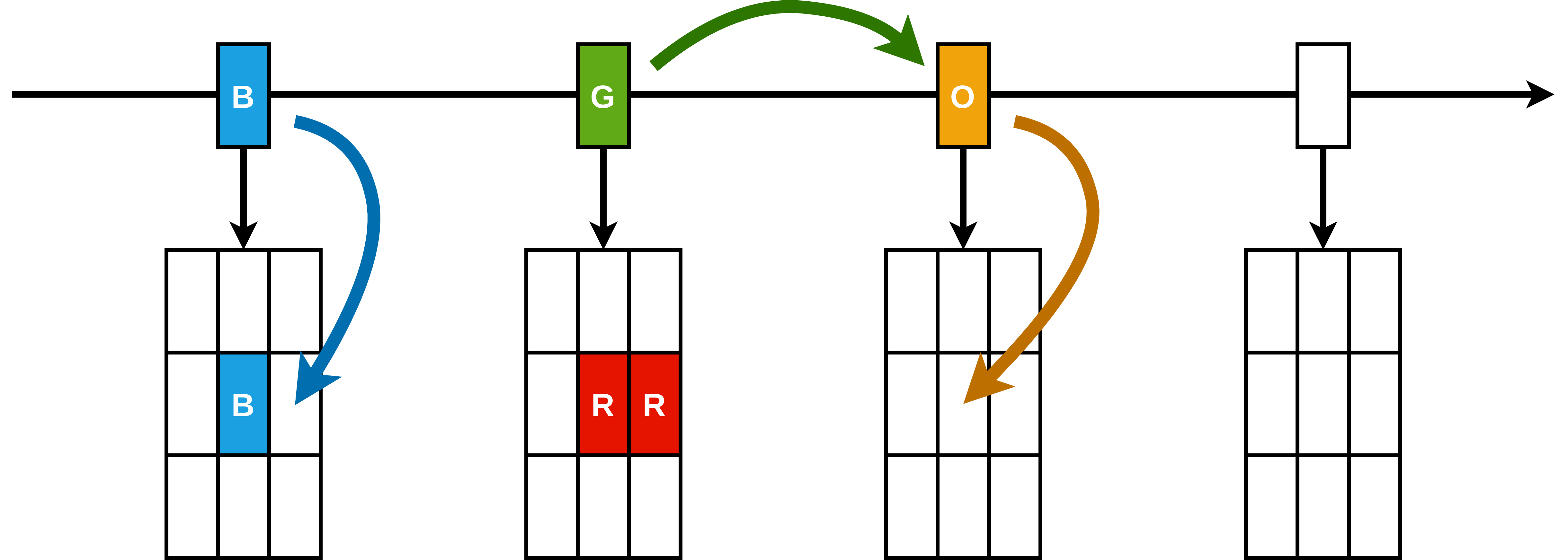}
    \caption{Sketch of the matrix filling mechanism for non-isolated SPs. SPs with same colour (label) are neighbours with active pixels. The blue SP (B) fills the first matrix in the line that is already populated with one of its neighbours. The green SP (G) does not belong to any of the already populated matrices, so it moves forward. The orange SP (O) has reached an non-initialised matrix, so it fills the centre.}
    \label{fig:matrix-filling}
\end{figure}

\begin{figure}[tb]
    \centering
    \includegraphics[width=20pc]{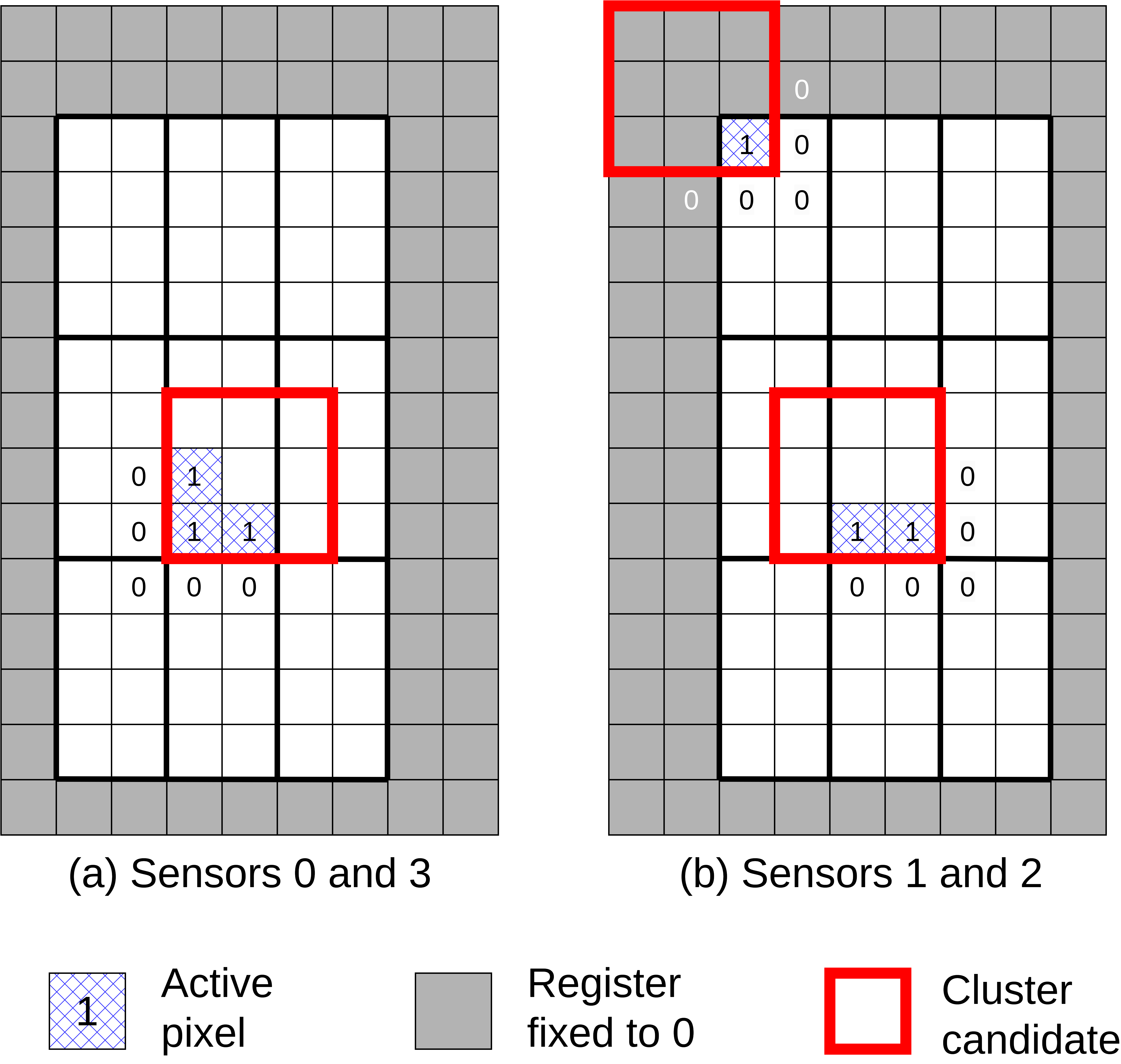}
    \caption{Matrix edges and pattern orientations (a) for sensors 0 and 3 (b) for sensors 1 and 2.}
    \label{fig:edges}
\end{figure}

This highly parallelised mechanism is key to the successful operation of the architecture at the extremely high throughput levels required by the LHC collision rate. However, the amount of FPGA logic resources needed to implement a complete bit-matrix map of the approximately $40$ millions pixels of the VELO detector would be excessive. This corresponds to approximately 780 thousand pixels in input to each VELO readout card.
To overcome this problem, a sparse representation of the bit matrix is adopted, breaking it down into a set of small matrices of fixed size, that get dynamically allocated only for the regions of the detector that contain active pixels. After some optimisation studies, accounting for  a maximum detector occupancy of around 0.125\%~\cite{velo-upgrade}, an average size of reconstructed clusters of 2 pixels,
and computational requirements, the size of each small matrix has been chosen to cover the same area as 3$\times$3 SPs, that is, 6$\times$12 individual pixels (Fig.~\ref{fig:matrix-filling}).
In order to reconstruct cluster candidates having an anchor pixel lying near the edge, each matrix is surrounded by edges of registers fixed at zero, as shown in Fig.~\ref{fig:edges}. These edges are not used during the filling process, but are necessary to determine the 3$\times$3 cluster candidate when there are active pixels at the edge of the 3$\times$3 matrix. An example of such a configuration is shown in Fig.~\ref{fig:edges}b.
The width of the edges is determined by the VELO sensor number, the allowed patterns, and the cluster candidate topology (Fig.~\ref{fig:pixel-patterns}).

To allow the allocation of matrices to proceed in real time without any delay, matrices are organised in a sequential chain for each VELO sensor, with SP data flowing continuously along the chain at the same rate as they are fed into the clustering block.
All matrices are initialised as empty. When a SP arrives at an empty matrix, it fills the centre of the matrix and it defines the physical location of the matrix inside the VELO detector, as well as the set of coordinates of the other SPs that can fill it. The allocated position of the matrix is checked against the coordinates of every SP going through the chain. If a SP belongs to the region inside the matrix, it is used to fill its appropriate location, otherwise it moves forward along the input line. Eventually, every SP gets stored in some matrix of the chain. An explanatory diagram illustrating the mechanism is shown in Fig.~\ref{fig:matrix-filling}.

When the input flow of SPs has ended, data from each matrix are copied in a single clock cycle to a twin matrix (pattern recognition matrix) where cluster finding is performed. In this way, the input matrix is ready to accept data from the next event immediately. The pattern recognition of all potential cluster candidates in this twin matrix is then performed, and the local coordinates of the centroid, with respect to the anchor pixel position, of each found cluster are determined using a LUT. The absolute position of the cluster candidate is obtained as the sum of three vectors of coordinates: the position of the matrix with respect to the sensor, the position of the anchor pixel with respect to the matrix, and the position of the reconstructed cluster with respect to the anchor pixel.

The clustering algorithm described has several parameters that can be tuned to optimise its performance in terms of speed, efficiency and quality of the reconstruction. The shape and size of the matrix are determined by how non-isolated SPs are arranged together, whereas the distribution of the number of SPs with neighbours per event sets the number of matrices that need to be instantiated. 
The implementation of the above algorithm as FPGA firmware does not allow the number of matrices to be dynamically adjusted to cope with the variable number of non-isolated SPs per event. However, from LHCb simulations we determined that a fixed number of 20 matrices per VELO sensor is sufficient to ensure that less than 0.1\% of the SPs overrun the matrix chain at the nominal Run~3 instantaneous luminosity ($2 \times 10^{33}$~cm$^{-2}$s$^{-1}$), and this number was adopted for the final implementation.
SuperPixels exceeding this limit are not discarded. Instead, partial information is extracted from them, by resolving them via a LUT as if they were isolated. This approach avoids inefficiencies, at the expense of a slight increase in the number of split clusters, since more than one cluster ends up being reconstructed from a single group of neighbouring pixels, when they happen to be spread over multiple SPs~\cite{luca-thesis,giovanni-thesis}. 
These clusters are flagged in the output as ``non-isolated'', to allow the reconstruction algorithms in the HLT to deal with them properly.

The LHCb experiment foresees to collect data also for heavy-ion collisions~\cite{heavy-ion}. A modified version of the clustering architecture will be used to cope with the higher number of SPs (around six times larger than in $pp$ collisions). Due to the limited amount of FPGA resources, the same matrices will be used multiple times to accommodate all SPs. The cluster reconstruction of a heavy-ion event requires more time than the $pp$ case, given the higher number of SPs, however, due to the much lower interaction frequency (50~kHz) the firmware can easily provide the necessary throughput also in this case.

\section{Physics performance}
\label{sec:phys-perf}

The FPGA cluster-finding architecture was designed with the intent of replacing the raw pixel data with reconstructed hit coordinates at the detector readout level. Except when running in debug mode (that preserves the full original information together with the cluster data), the raw pixel data are discarded and can not be recovered at any later stages.    
Therefore, extensive simulation studies were performed to assess the effect of using the FPGA-reconstructed clusters on the physics performance of the LHCb reconstruction, both at the HLT1 and HLT2 stages. This was compared to the alternative scenario in which the VELO hits are reconstructed by a full-fledged software reconstruction within the HLT system, free from all the constraints imposed by the FPGA architecture, and from the severe throughput requirements of operating at pre-build level (30\mhz vs. 0.17\mhz, where a farm of about 170 GPUs is assumed for HLT1).

For the sake of generality, comparisons are made to a CPU-based clustering algorithm, that is free from implementation-specific constraints. However, the actual HLT1 implementation at LHCb is GPU-based, but its performance is indistinguishable from the CPU version we take as reference.
The key differences between the FPGA and CPU algorithms that can potentially affect reconstruction performances are the cluster finding mechanism, the maximum cluster size in the FPGA algorithm (limited to a 3$\times$3 pixel grid), and the constraints of the FPGA matrix filling scheme. They can potentially lead to inefficiencies, cluster splitting, or incomplete reconstruction of some clusters.
\begin{figure}[tb]
    \centering
    \includegraphics[width=18pc]{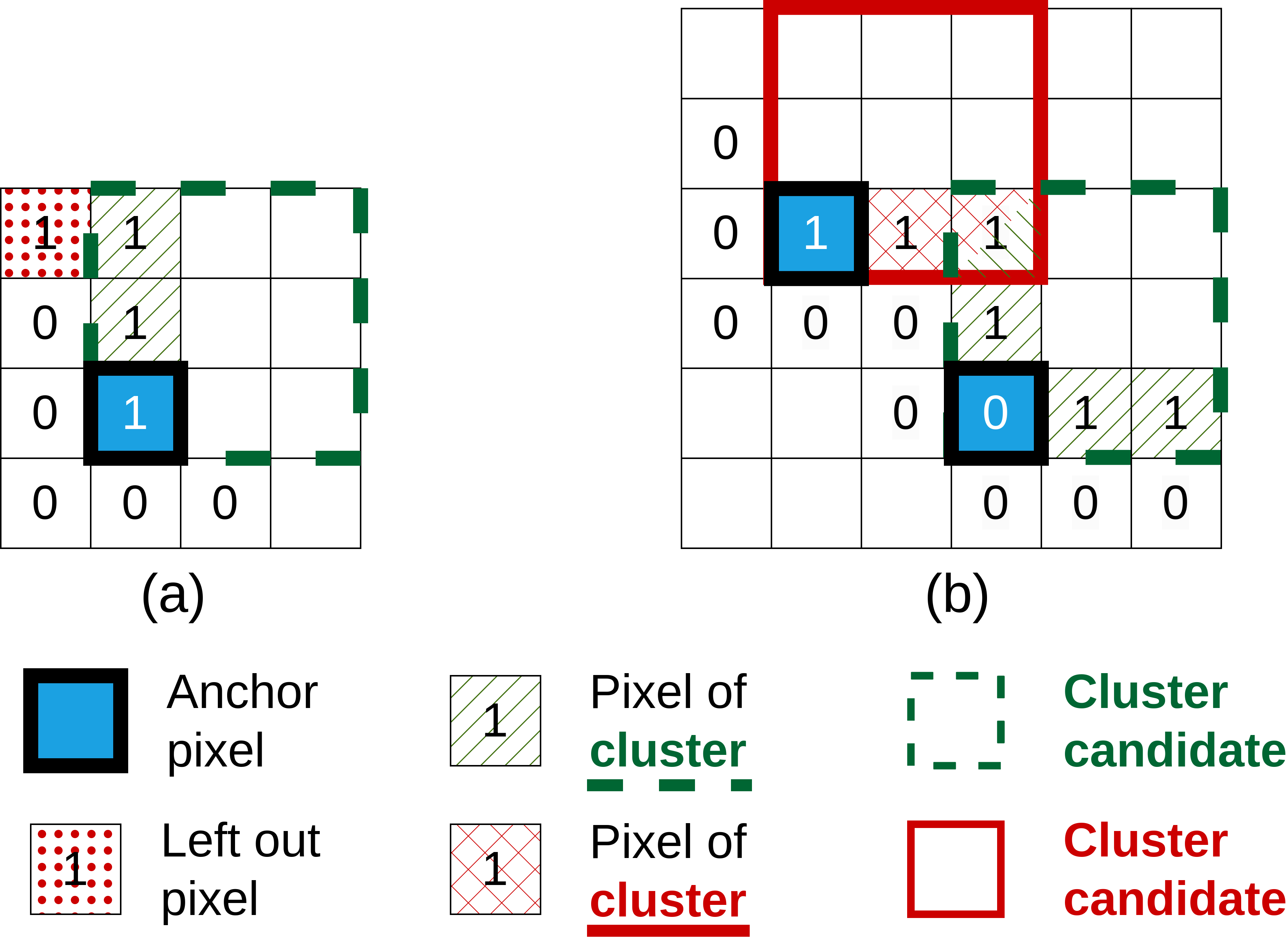}
    \caption{Example of corner cases of the FPGA clustering algorithm: (a) partial cluster reconstruction and (b) cluster splitting.}
    \label{fig:peculiar-behavior}
\end{figure}
An example of partial cluster reconstruction is illustrated in Fig.~\ref{fig:peculiar-behavior}a, where the red pixel is left out of the reconstructed cluster. The shift of the reconstructed hit position may lead to a degradation of the precision on the reconstruction of the particle trajectory, or even to a loss of efficiency if the associated track is not reconstructed at all. Figure~\ref{fig:peculiar-behavior}b shows an example of cluster splitting, where the algorithm finds two clusters, with a pixel in common, from six contiguous active pixels. 

\begin{figure}[tb]
    \centering
    \includegraphics[width=20pc]{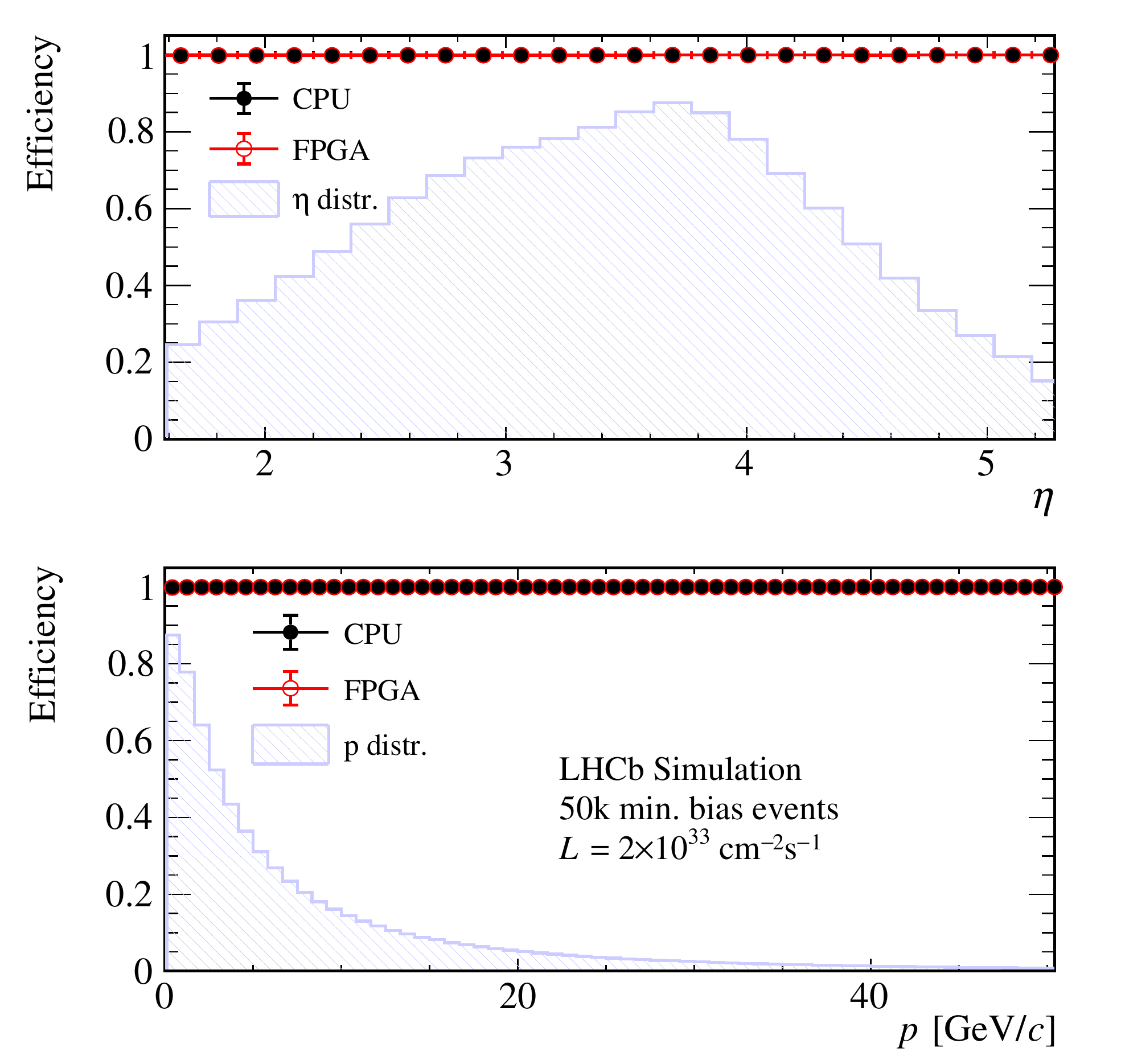}
    \vspace*{-0.3cm}
    \caption{Efficiency in reconstructing clusters as a function of (top) the pseudo-rapidity and (bottom) the momentum of the associated tracks for the CPU- and FPGA-based algorithm.  Clusters, from tracks that can be reconstructed using only information from VELO hits, are shown. The blue histograms show (top) the pseudo-rapidity and (bottom) the momentum distributions of the tracks.}
    \label{fig:cluster_eff}
\end{figure}

To perform the studies, a faithful software simulation of the FPGA-based clustering algorithm has been produced. This simulation was integrated within the official LHCb software simulation, and a CPU--FPGA comparison was performed on a sample of 50,000 bunch crossings, each containing an average number of $7.6$ $pp$ interactions, at an instantaneous luminosity of $2\times10^{33}~\,\mathrm{cm}^{-2}\,\mathrm{s}^{-1}$. This corresponds to a total number of about $10^8$ SPs ($7\times10^7$ clusters), generated at the foreseen LHCb-Upgrade running conditions with a centre of mass energy of 14\tev.
The efficiency in reconstructing VELO clusters is defined as the ratio between the number of Monte Carlo (MC) hits matched to a cluster and the total number of MC hits. Only MC hits that produce enough charge in the detector to activate at least one pixel are considered. A MC hit is matched to a cluster if they share at least one pixel. The efficiency in reconstructing VELO clusters of the FPGA-based algorithm is about 99.8\%, and almost indistinguishable from that of the CPU algorithm, as illustrated in  Fig.~\ref{fig:cluster_eff}. Here, the efficiency of reconstructing clusters from tracks that can be reconstructed using information from VELO hits only is shown as a function of the pseudo-rapidity ($\eta$) and momentum of the tracks.
The overall FPGA cluster inefficiency, with respect to the CPU algorithm,  is below 0.1\% within the LHCb geometrical acceptance ($2<\eta<5$).

\begin{figure}[tb]
    \centering
    \includegraphics[width=20pc]{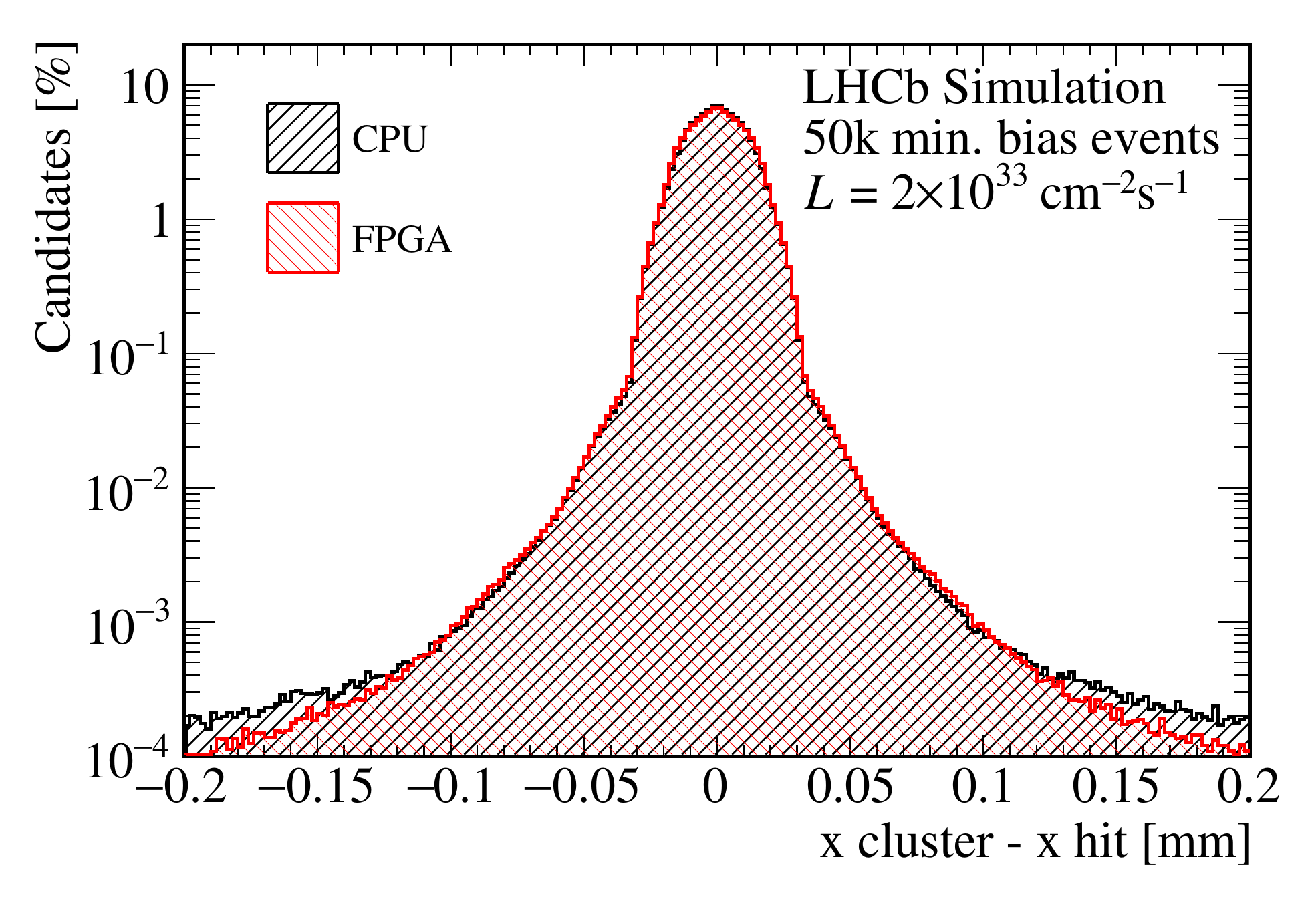}
    \vspace*{-0.3cm}
    \caption{Distributions of cluster residuals, along the $x$ coordinate, for the CPU and FPGA based clustering algorithms. Distributions are normalized to unity. Similar results are obtained for the y coordinate.}
    \label{fig:cluster_res}
\end{figure}

The quality of the reconstructed clusters is also studied by looking at the distributions of cluster residuals. The residual is defined as the distance between the position of the reconstructed cluster and the true coordinates of the hit generated by the passage of the particle on the  associated detector layer. A comparison between residual distributions of reconstructed clusters, between the CPU and FPGA algorithm, is shown in Fig.~\ref{fig:cluster_res}. Distributions are plotted over the $x$ coordinate, in the LHCb global reference frame.
The two distributions are indistinguishable in the core, with very small differences in the tails. It is also checked that most of the non-reconstructed hits are of inferior quality, sitting in the tails of the resolution curve.

Extensive studies are also performed to measure the quality of the full track reconstruction, when FPGA VELO clusters are used.
The trajectories of charged particles traversing the tracking system are reconstructed from hits in some of the three tracking detectors, that is the VELO, the Upstream Tracker (UT) placed upstream of the magnet, and the Scintillating Fibre (SciFi) detector placed downstream of the magnet~\cite{tracker-upgrade}. 
Tracks reconstructed using only hits from the VELO detector are called VELO tracks. VELO tracks having $\eta < 2$ are used only for the primary vertex reconstruction while those with $2 < \eta < 5$ can be extended in the forward region to attach hits from the SciFi detector, and optionally from the UT. These tracks are called ``long tracks''. As they traverse the whole magnetic field of the LHCb detector, they have the most precise measurement of the momentum and therefore are key for physics analyses.
Table~\ref{tab:phys-performance} shows a comparison between the CPU- and FPGA-based reconstruction performances for VELO tracks and for the VELO segment of long tracks.
It also reports the relative fraction of clone-reconstructed tracks with respect to the total number of tracks in the category they belong to, and the relative fraction of ghost-reconstructed tracks with respect to the total number of tracks. A clone is defined as any additional reconstructed track matching an already truth-matched Monte Carlo track, whereas a ghost is a reconstructed track not associated with any true Monte Carlo track\cite{tracking-defs}.
\begin{table}[tb]
\centering
\caption{VELO tracking efficiency, relative fraction of clone and ghost tracks, comparing CPU- and FPGA-based clusters.}
\label{tab:phys-performance}
\setlength\tabcolsep{3.5pt}
\begin{tabular}{cccc} 
\hline
\\[-1.8ex] 
\multicolumn{1}{c}{Track type} & \multicolumn{1}{c}{Quantity} & \multicolumn{1}{c}{CPU clusters [\%]} & \multicolumn{1}{c}{FPGA clusters [\%]} \\[2pt]
    \hline 
    \rule{0pt}{3ex}
    \multirow{2}{*}{All VELO tracks} & efficiency & 98.254 $\pm$ 0.007 & 98.254 $\pm$ 0.007 \\
    & clone & \, 1.231 $\pm$ 0.006 & \, 1.234 $\pm$ 0.006\\[2pt]
    \hline
    \rule{0pt}{3ex}
    \multirow{2}{*}{Long tracks} & efficiency & 99.252 $\pm$ 0.006 & 99.252 $\pm$ 0.006 \\
    & clone & \, 0.806 $\pm$ 0.006 & \, 0.806 $\pm$ 0.006\\[2pt]
    \hline
    \rule{0pt}{3ex}
     &ghost & \, 0.848 $\pm$ 0.003 & \, 0.928 $\pm$ 0.003\\[2pt]
     \hline
\end{tabular}
\vspace{4pt}
\end{table}
The efficiencies and clone fractions are almost indistinguishable when comparing CPU and FPGA algorithms for VELO and long tracks, not displaying any perceptible systematic difference.
The fractions of ghost tracks differ at the per-mille level. This difference is due to tracks in the pseudorapidity region below 1.5. These tracks graze VELO sensors at a very low angle, and produce very large clusters. For this reason, the position of the particle hitting the detector and creating the cluster is unlikely to be accurately measured regardless of the clustering algorithm.

\begin{figure}[tb]
    \centering
    \includegraphics[width=20pc]{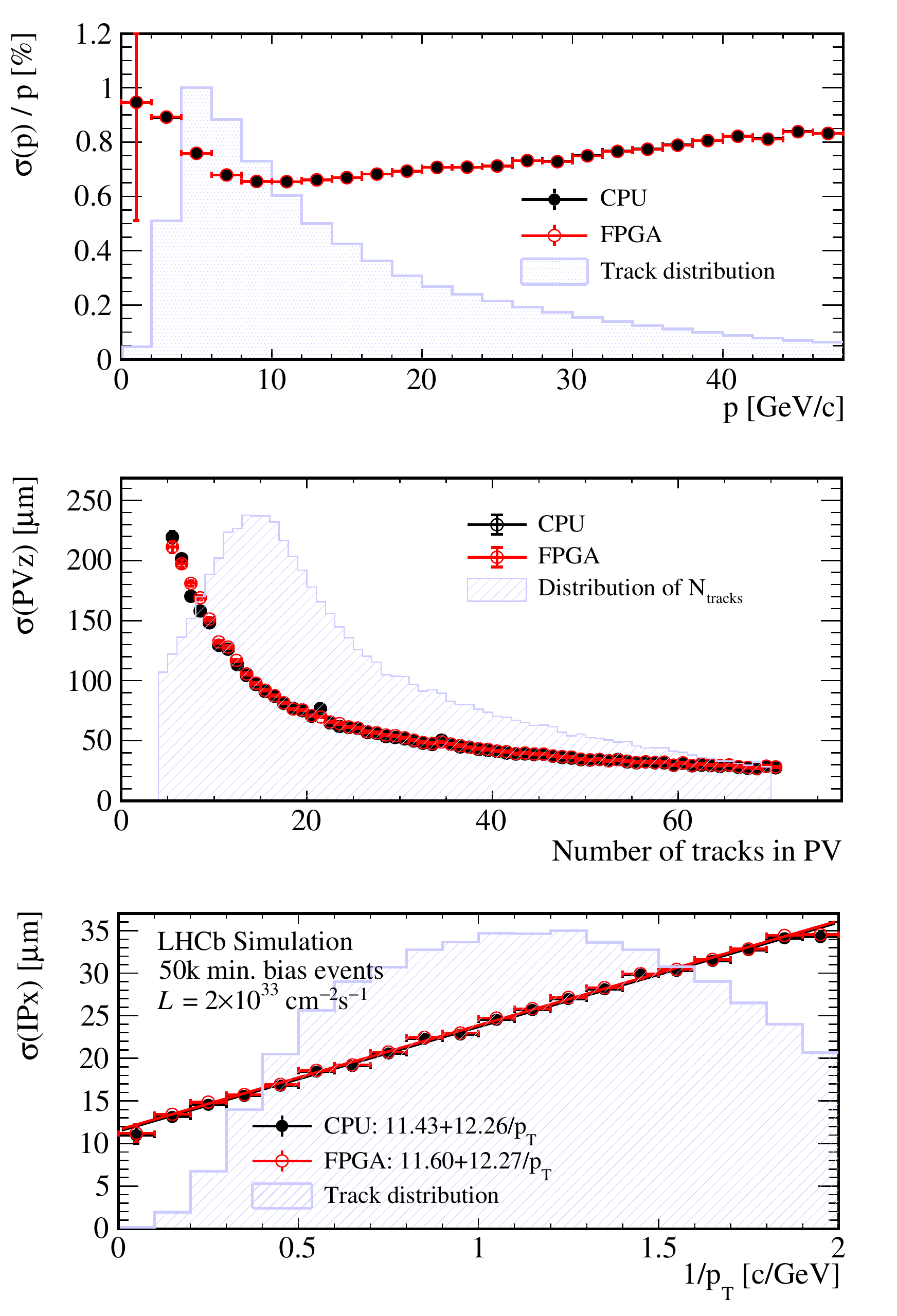}
    \caption{Track reconstruction resolutions for the CPU and FPGA based clustering algorithms: (top) momentum resolution as a function of the momentum, (middle) primary vertex resolution along the beam direction as a function of the number of tracks in the reconstructed primary vertex and (bottom) impact parameter resolution along the horizontal direction as a function of the inverse of the transverse momentum. Impact parameter resolutions are fitted with a linear function. The blue histograms show the distributions of the (top) momentum of the reconstructed tracks, (middle) number of reconstructed tracks per primary vertex and (bottom) inverse of the transverse momentum of the reconstructed tracks.}
    \label{fig:itrack_res}
\end{figure}

The quality of the reconstructed tracks is also studied in terms of momentum, primary vertex and impact parameter resolutions, as shown in Fig.~\ref{fig:itrack_res}.
The robustness of the algorithm is also verified against occupancy and relative fraction of large clusters~\cite{giovanni-thesis}. Differences in reconstruction quality between the FPGA and the CPU implementations do not show any trend as a function of these probes.
In conclusion, all studies have shown that FPGA-reconstructed clusters lead to a quality of track reconstruction that is effectively indistinguishable from the software reconstruction.

\section{Implementation details and integration}
\label{sec:implementation}

\begin{figure*}[tb]
    \centering
    \includegraphics[width=0.9\linewidth]{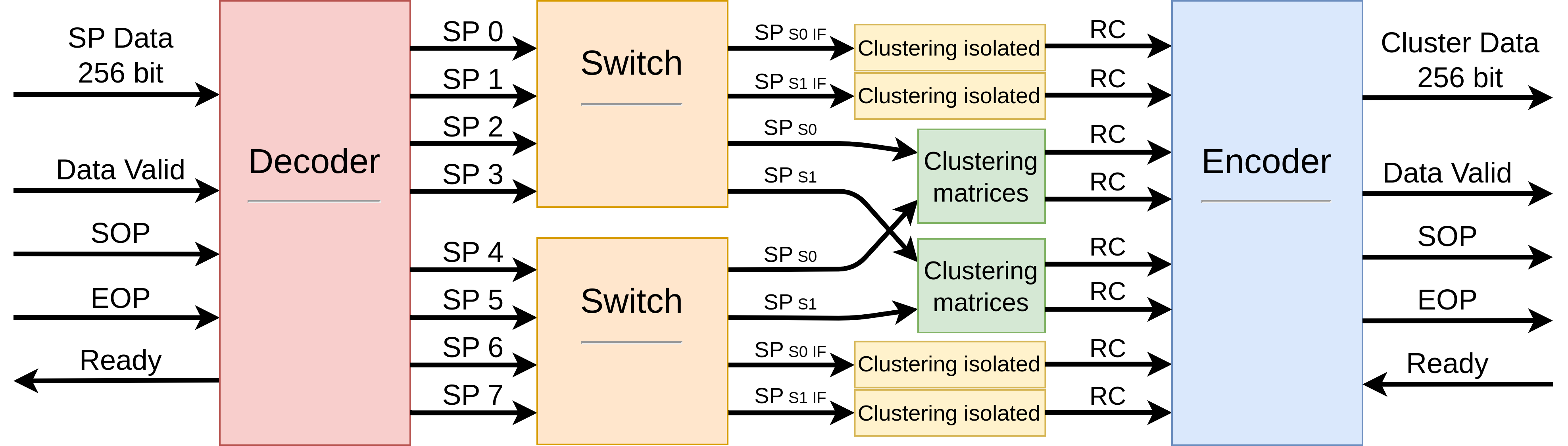}
    \caption{Basic blocks of the clustering architecture. VELO data are received as 256-bit words, each containing 8 SPs. A ``Data Valid'' signal states whether the incoming data are valid. Start of package (SOP) and end of package (EOP) signals delimit the start and the end of the data corresponding to each event. The clustering block sends a ready signal to the previous architecture component when it is ready to accept data. The ``decoder and isolation flagging'' splits the 256-bit bus into eight 32-bit wide busses, each containing one SP. It also flags SPs that do not have any active neighbour SPs (isolation flag). A pair of switches arrange SPs by sensor (S0/S1) and by isolation flag (IF). The ``clustering isolated'' and ``clustering matrices'' blocks reconstruct clusters, that are encoded back into 256-bit words by means of an encoder.}
    \label{fig:firmware-structure}
\end{figure*}

Given the indistinguishable performance of the FPGA-based clustering algorithm with respect to the software-based one, the LHCb collaboration decided to integrate the cluster finder architecture within the TELL40 cards that perform the readout of the VELO, exploiting spare FPGA resources not utilised by the readout firmware. The VELO time-ordering firmware, plus the common LHCb firmware, takes up about 44\% of its logic resources and 64\% of its M20K memory blocks.  

Each VELO TELL40 receives data from a single VELO module. Two independent and identical parallel processing chains are implemented in the FPGAs, each of which receives and processes data from one VELO half-module~\cite{velo-firmware}. The clustering architecture, with all needed ancillary logic, is integrated as a self-contained block at the end of each chain, and it has, therefore, two identical instances running in parallel, in analogy with the readout firmware (Fig.~\ref{fig:firmware-structure}). The output of the clustering is transmitted out of the readout card through its PCIe interface to the host server, which assembles the data from different subdetectors for each event. 

The clustering architecture is itself composed of several units, each devoted to a specific task (Fig.~\ref{fig:firmware-structure}).
Firstly, a decoding and flagging stage splits data into separate streams, while flagging isolated SPs. Secondly, a pair of switch blocks sends data to the cluster processing blocks. Reconstructed clusters are then finally encoded with the chosen output format. 
A back-pressure mechanism is implemented throughout the pipeline: each processing block sends a ``ready'' signal to the previous unit when it is capable of receiving data.

A detailed description of the firmware implementation, its integration and commissioning within the LHCb data acquisition, can be found in Ref.~\cite{giovanni-thesis}.

\subsection{Clock domains}

Each unit in an instance of the clustering architecture writes its output to a FIFO that is read by the subsequent unit. The purpose of the FIFOs is twofold. Firstly, they allow buffering and flow control between the clustering units. In addition, FIFOs allow data synchronisation between different clock domains. In our application, the decoder and the encoder stages run on a 250\mhz clock, whereas the switch and clustering processing block use a 350\mhz clock. These values have been chosen to ensure that the system as a whole can provide a throughput in excess of 30\mhz (see Sect.~\ref{sec:resource_throughput}), while still respecting the timing constraints due to internal signal propagation in all of its parts.

\subsection{Data formats}

Active SPs are encoded as 32-bit words. Each word contains the pixel hitmap (8 bits), the SP position inside the sensor (15 bits) and the sensor identifier within the sensor pair (1 bit).
Each VELO sensor is made of 256$\times$768 pixels. Each SP is composed of 4$\times$2 pixels, such that 6 bits are needed to specify the SP row whereas 9 bits are required for the column. One extra bit is needed to identify the source sensor, as each data chain receives SPs from two sensors.

\begin{figure}[tb]
    \centering
    \includegraphics[width=1\linewidth]{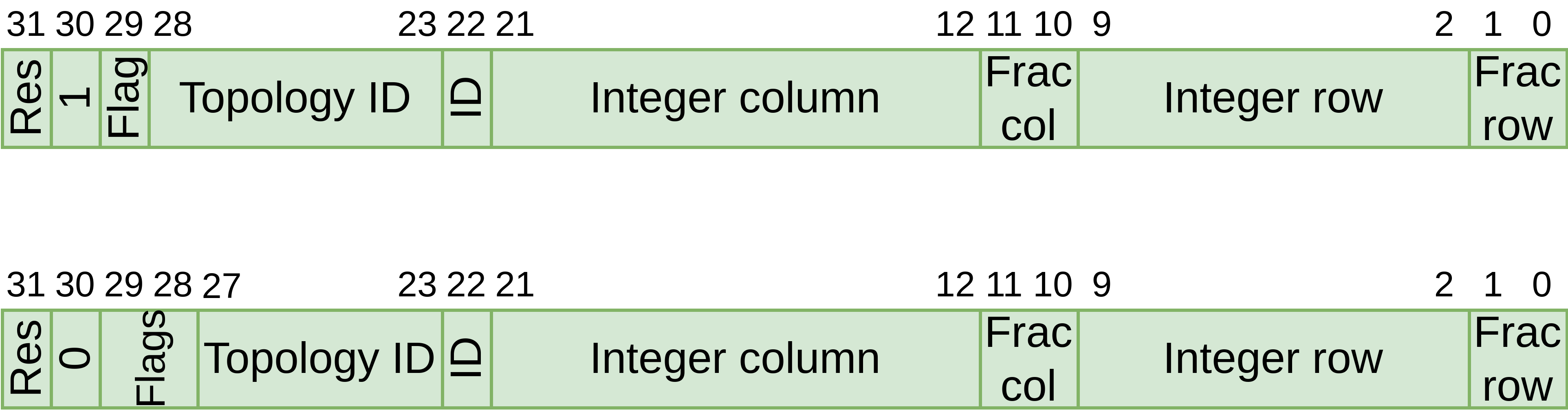}
    \caption{Data formats for (top) clusters reconstructed from isolated SPs and (bottom) clusters reconstructed from not isolated SPs. Bit 31 (Res) is reserved for internal use.}
    \label{fig:cluster-data-format}
\end{figure}

Clusters are encoded in 32-bit output words, as sketched in Fig.~\ref{fig:cluster-data-format}. 
Of these, 22 bits are used to specify the position of the cluster centroid, with 18 bits specifying the position of the pixel where the cluster centroid is located (Integer column and Integer row), and additional 4 bits are used to specify the position of the centroid within the pixel, in units of $1/4$ of a pixel (Frac col and Frac row). Analogously to SP data, 1 bit is used to identify the sensor (ID). Eight additional bits are used to encode a cluster-topology identifier (Topology~ID) and the reconstruction-quality flags (Flags). The topology identifier is used to distinguish cluster topologies that share the same centroid position within the pixel, so that the full cluster topology can be retrieved. If the cluster is reconstructed from an isolated SP (bit 30 = 1 in Fig.~\ref{fig:cluster-data-format} top), six bits are used to store the topology identifier, whereas five bits are needed to store the identifier for clusters reconstructed through the matrices (bit 30 = 0 in Fig.~\ref{fig:cluster-data-format} bottom). 
The cluster topology information is used both for monitoring purposes and for the ultimate optimisation of tracking performance, as the uncertainty associated with the 2D position of a cluster depends on its topology. The reconstruction-quality flags allow to distinguish between: clusters from isolated SPs, clusters reconstructed inside a matrix, and clusters built from SPs overflowing the maximum number of instantiated matrices (which are arbitrarily treated as isolated). For clusters reconstructed within matrices, the word contains two additional quality flags, which specify whether a cluster was fully contained in the $3\times3$ grid, and whether the grid touched the boundary of the host matrix (which potentially means that the reconstructed cluster is a fragment of a larger cluster).

\subsection{Input-output interfaces}

Our architecture block requires a ``valid'' signal to confirm the validity of the current input data word. Additional start of package (SOP) and end of package (EOP) signals allow to separate data coming from different events. The SOP signal is received with the first word of every event, whereas the EOP is generated together with the last input word.
The ``valid'', SOP and EOP signals are also present on the output side, where clusters are transmitted.

\subsection{Decoder block}

Input data to the clustering firmware arrive grouped in 256-bit words, each carrying 8 SP words. The first block is a decoder, which splits the 256-bit words into eight 32-bit streams.
The decoder is also responsible for converting the SOP-EOP protocol to the EndEvent (EE) protocol used within the clustering architecture: a 32-bit EE word is interposed between SPs of different events in all the 8 data streams. Each EE word carries a specific flag to distinguish it from SPs, and an event identifier (5 bits), that can be used during subsequent data processing to cross-check data synchronisation.

\subsection{Isolation flagging}

\begin{figure}[tb]
    \centering
    \includegraphics[width=1\linewidth]{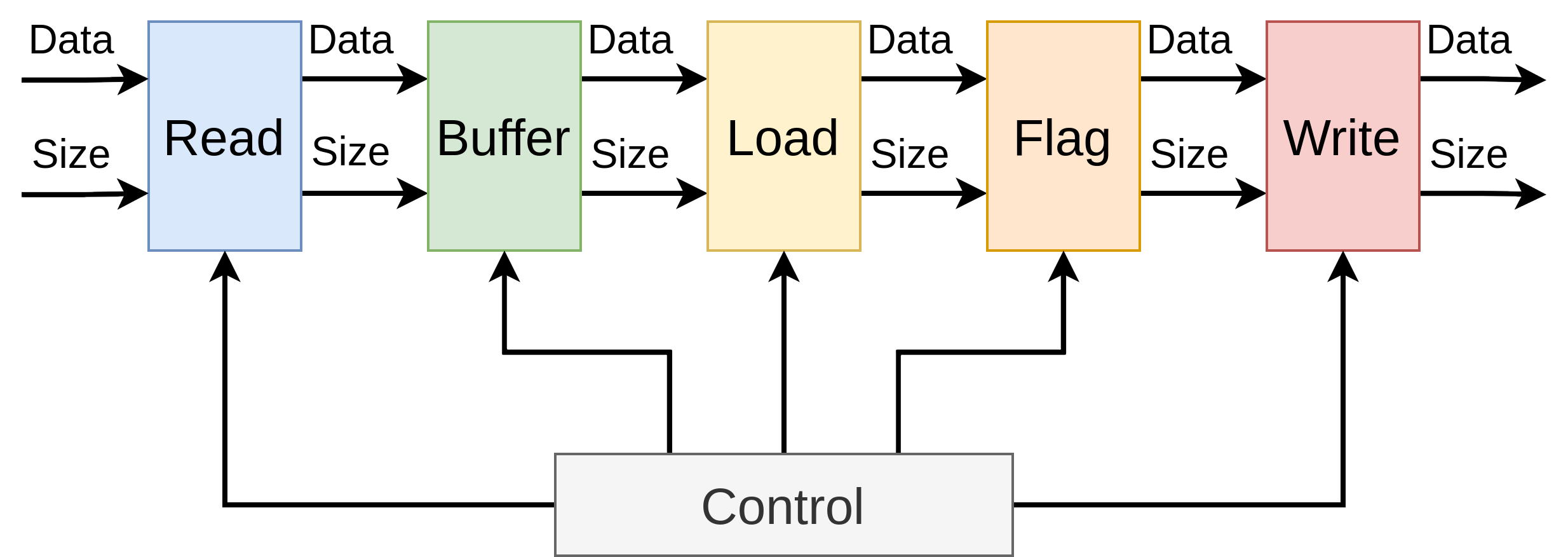}
    \caption{Block diagram of the isolation flagging.}
    \label{fig:icf-flagging}
\end{figure}

Within the decoder block, SPs are flagged with an isolation bit. The flagging process includes five steps: read, buffer, load, flag and write, arranged in a pipeline, as shown in Fig.~\ref{fig:icf-flagging}.

First, all SPs of a given event are read and stored into registers. The maximum number of SPs that can be stored in the read registers is not dynamically adjustable. It has been set to 144 based on the distribution of the expected number of SPs in the most crowded VELO module, requiring more than 98\% of the LHCb simulated events to be accommodated into the read registers.
In events where the number of SPs exceeds the size of the read registers, SPs are not sent to the flagging process but are instead bypassed and sent directly to the matrix chains. This causes a local slowdown of the entire chain as the matrices need to reconstruct clusters from a high number of SPs. This effect is included in the measurement of the average throughput of the entire system. In addition, as the number of input SPs increases, the fraction of SPs overflowing the number of available matrices gets higher, increasing the number of split clusters. The corresponding increase in split clusters is also taken into account in the evaluation of the reconstruction performance.

As soon as all SPs of one event have been received, the content of the read registers is copied to the buffer. This data exchange decouples the reading and flagging operations, allowing SPs of one event to be read in while the flagging of the previous event is still ongoing. The flagging process compares the coordinates of each SP to the ones of the other SPs in the same event. A status vector is used to store the isolation flag for each SP: if two SPs are found to be neighbours, the corresponding bits in the status vector are set to 1. SP comparisons are not all performed in a single clock cycle. On each clock, the load block extracts two subsets of 16 SPs each from the buffer (Fig.~\ref{fig:icf-flagging}). For each SP in the two subsets it also computes the set of coordinates to be matched by the neighbours by one-unit additions and subtractions of the coordinates of the SP row and column. The two SP subsets, together with the coordinates of the neighbours, are passed to the flag block that performs the $16\times 16$ comparisons on the two subsets. For each SP of the first subset, the flag block checks if the SP row is equal to one of the rows of the SPs in the second subset or to the row above or below; the same check is performed on columns. If both the row and the column checks yield a positive result, the two SPs are flagged as neighbours, and the corresponding bits in the status vector are set to 1. On each clock cycle, the load block selects a different pair of SP subsets from the buffer sending them to the flag block, until all possible combinations of 16-SP subsets have been checked. The described architecture allows reusing the same logic resources while updating the SP subsets to be flagged at each clock cycle. To perform the comparisons between $n$ 16-SP blocks, $n(n+1)/2$ clock cycles are needed.

The number of parallel comparisons performed for every clock cycle is the result of a trade-off between resource usage and throughput, and is based on the constraints of its use within the LHCb experiment.

As soon as all comparisons have been completed, the contents of the flagging registers and the status vector are copied to the write block, thus decoupling the flag and write processes. 
The write block is responsible for adding the isolation flag to the SP words and for sending flagged data to the next component, the switch. The data exchange within the read-buffer-load-flag-write pipeline is regulated by back-pressure: if a component cannot accept the data of an event because it is still processing the previous event, the control unit keeps the previous component on hold.

The decoder block, including flagging and bypass, uses 7\% of the logic and 1\% of the M20K memories available in an Arria 10 FPGA.

\subsection{Switch block}

The cluster processing chain receives SPs from both sensors of a VELO half-module. 
The switch, placed after the decoder, arranges SPs by sensor and by isolation flag, feeding them to the appropriate cluster-reconstructing blocks. 
Each of the two switching units shown in Fig.~\ref{fig:firmware-structure} performs a $4\to 4$ switching, allowing every input data word to be directed to any of the four output streams according to its flags, regardless of the origin input stream.
\begin{figure}[tb]
    \centering
    \includegraphics[width=1\linewidth]{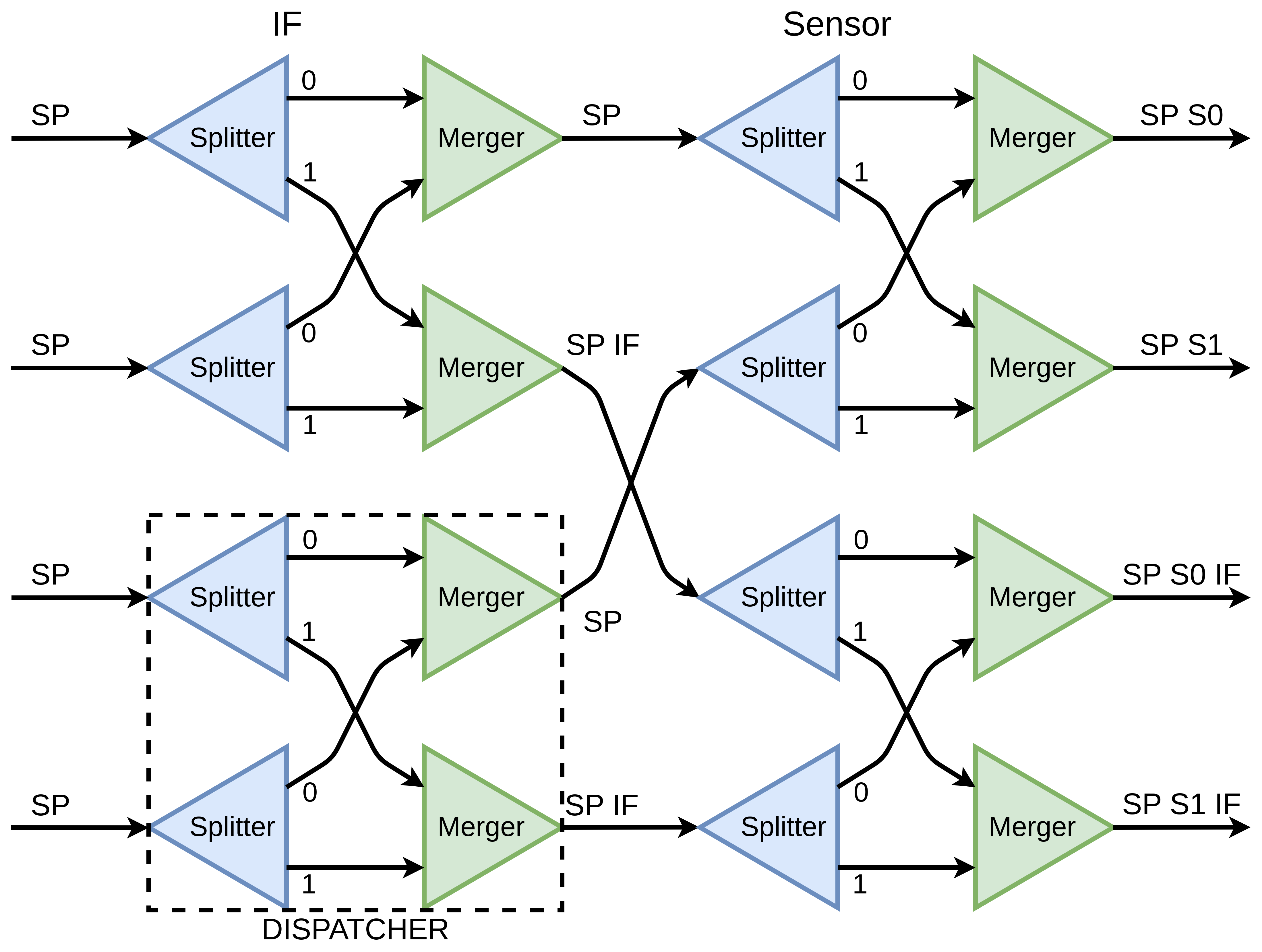}
    \caption{Block diagram of a 4 to 4 switching unit.}
    \label{fig:switch-block}
\end{figure}
The basic switch constituents are the splitter and the merger (Fig.~\ref{fig:switch-block}). The former has one input and two outputs, and it sends input data to one of the two outputs according to their isolation flag or origin sensor. The latter has two inputs and one output, and it routes two inputs in a single output line. Two splitters and two mergers combine to form a $2\to 2$ dispatcher.

\begin{figure}[tb]
    \centering
    \includegraphics[width=20pc]{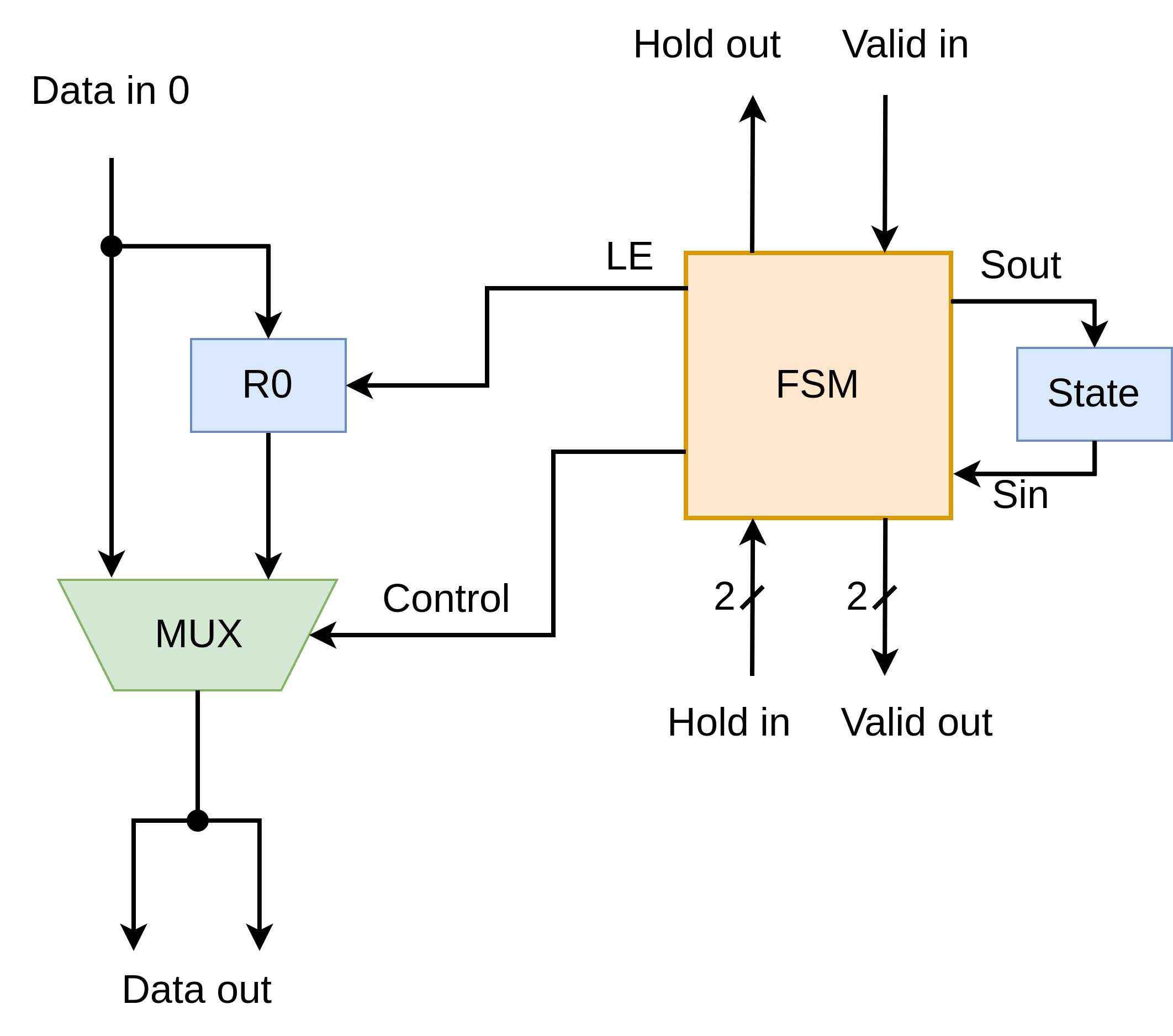}
    \caption{Splitter block diagram. R0 and State are registers, MUX is a multiplexer and FSM is a finite state machine that manages hold, valid, control and latch enable (LE) write signals.}
    \label{fig:splitter}
\end{figure}

The block diagram of the splitter is shown in Fig.~\ref{fig:splitter}. The splitter is based on a finite-state machine (FSM).
 The next state is determined by the R0 register state, the arrival of valid input data and the hold state of the following processing block.
On the arrival of valid input data, the FSM decides between sending it directly to the output and storing it in the R0 register, based on the input hold signal. In the latter case, a latch enable (LE) write signal is sent to the register. A multiplexer controlled by the FSM routes data to the output. 
If a SP is received then one of the two valid signals is set to 1, according to the routing scheme (isolation flag or sensor). If an EE signal arrives, it is sent to both outputs. The input hold signal determines whether data can be sent to the output. An output hold is generated as long as the R0 register is full, since no more data can be accepted as input, given the possibility of an input hold signal assertion.

\begin{figure}[tb]
    \centering
    \includegraphics[width=20pc]{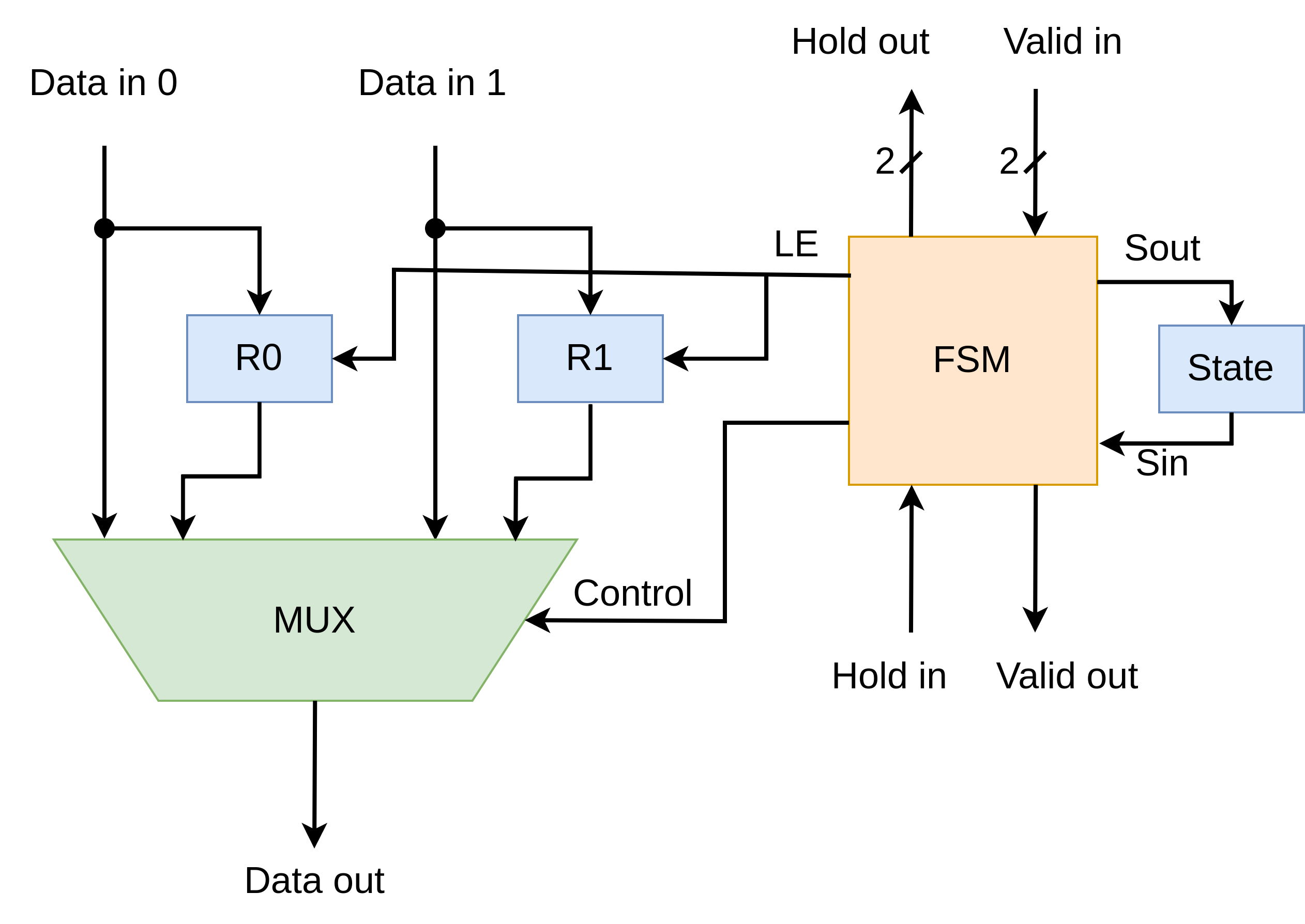}
    \caption{Merger block diagram. R0, R1 and State are registers, MUX is a multiplexer and FSM is a finite state machine that manages hold, valid, control and latch enable (LE) write signals.}
    \label{fig:merger}
\end{figure}

The block diagram of the merger is shown in Fig.~\ref{fig:merger}. As for the splitter, a FSM determines if input data can be sent directly to the output or must be stored in appropriate registers (R0 and R1).
If an EE word arrives on one of the inputs, it is stored until a second EE word arrives at the other input. The two EE words are then compared and, if their event IDs match, a single EE word is output; otherwise, a sync error signal is set to 1.

\subsection{Cluster reconstruction}

\begin{figure}[tb]
    \centering
    \includegraphics[width=20pc]{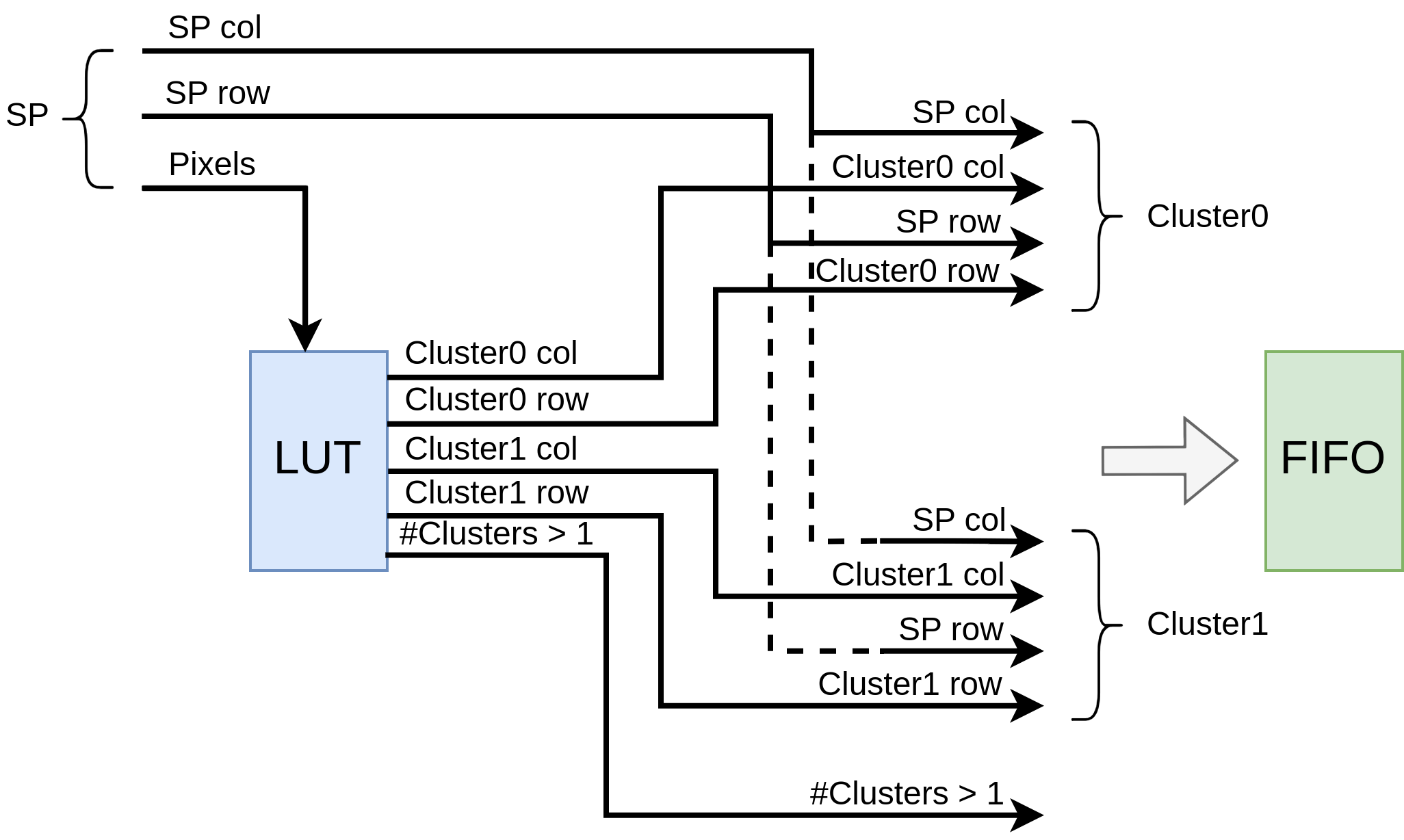}
    \caption{Cluster reconstruction of isolated SPs by means of a LUT.}
    \label{fig:isolated}
\end{figure}

All isolated SPs, identified by the switch, are sent to the corresponding clustering block, and are resolved by means of a LUT, as shown in Fig.~\ref{fig:isolated}. 

The LUT reconstructs the cluster centroid from the active-pixel hitmap extracted from the SP word. The cluster word is built by combining the LUT output with the original SP row and column. If two different clusters are reconstructed inside an isolated SP, a bit is set to 1, and the two outputs are combined by a merger block (Fig.~\ref{fig:merger}). Reconstructed clusters are then sent to the output FIFO.

\begin{figure}[tb]
    \centering
    \includegraphics[width=1\linewidth]{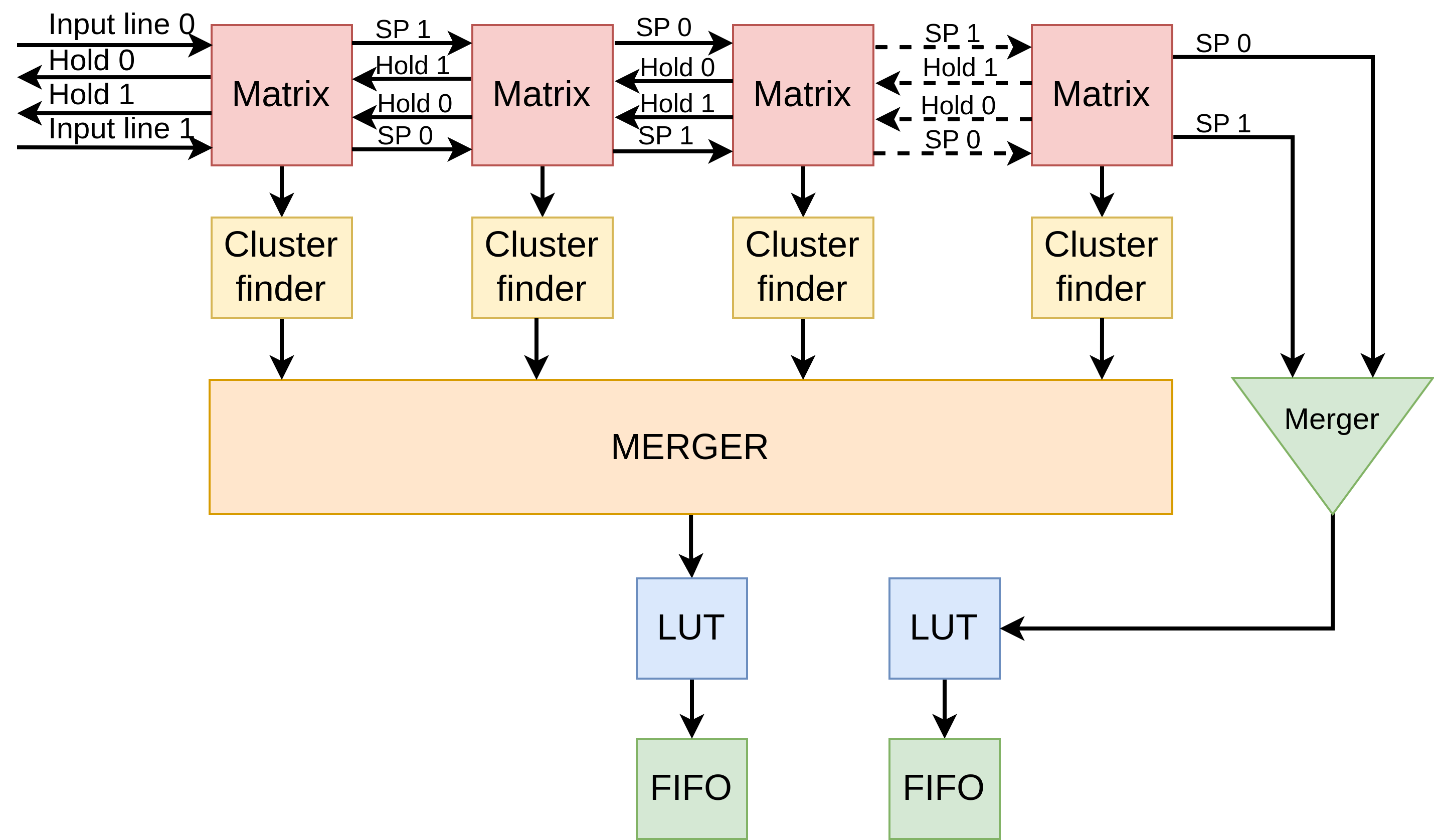}
    \caption{SP distribution in a matrix chain. Clusters are reconstructed through the cluster finder block and merged into a FIFO.}
    \label{fig:matrix-chain}
\end{figure}

The reconstruction of clusters from non-isolated SPs requires two different processing steps. Input data are first sent and distributed in the matrix chain, and then, when matrices have been filled with SPs, the actual reconstruction of the clusters takes place, as shown in Fig.~\ref{fig:matrix-chain}. 
In order to ensure a high throughput, each matrix receives data from two parallel input lines. Each input line is combined with a hold signal, that is propagated backwards through the whole chain to control the data flow by back-pressure to avoid data loss. As the first SP populates a matrix, a set of coordinates is calculated and stored, to be matched with all further SPs arriving at the same matrix.
The initialisation of an empty matrix is done using only one of the two input lines, since only a single SP can enter the centre of the matrix at a time. A second SP coming simultaneously from the other parallel line would need the coordinates of free slots to fill the matrix, that  cannot be immediately available for timing constraints.
For this reason, input line 0 (Fig.~\ref{fig:matrix-chain}) has the priority over input line 1, which is put on hold as the matrix is initialised. In order to keep a good load balancing, input lines are swapped when going from one matrix to the next:  line~0 of a matrix feeds line~1 of the next matrix and vice-versa.
When EE words have arrived on both input lines, the content of the matrix is moved to the cluster-finder block. An error is raised if two different EE signals are detected.

\begin{figure}[tb]
    \centering
    \includegraphics[width=0.93\linewidth]{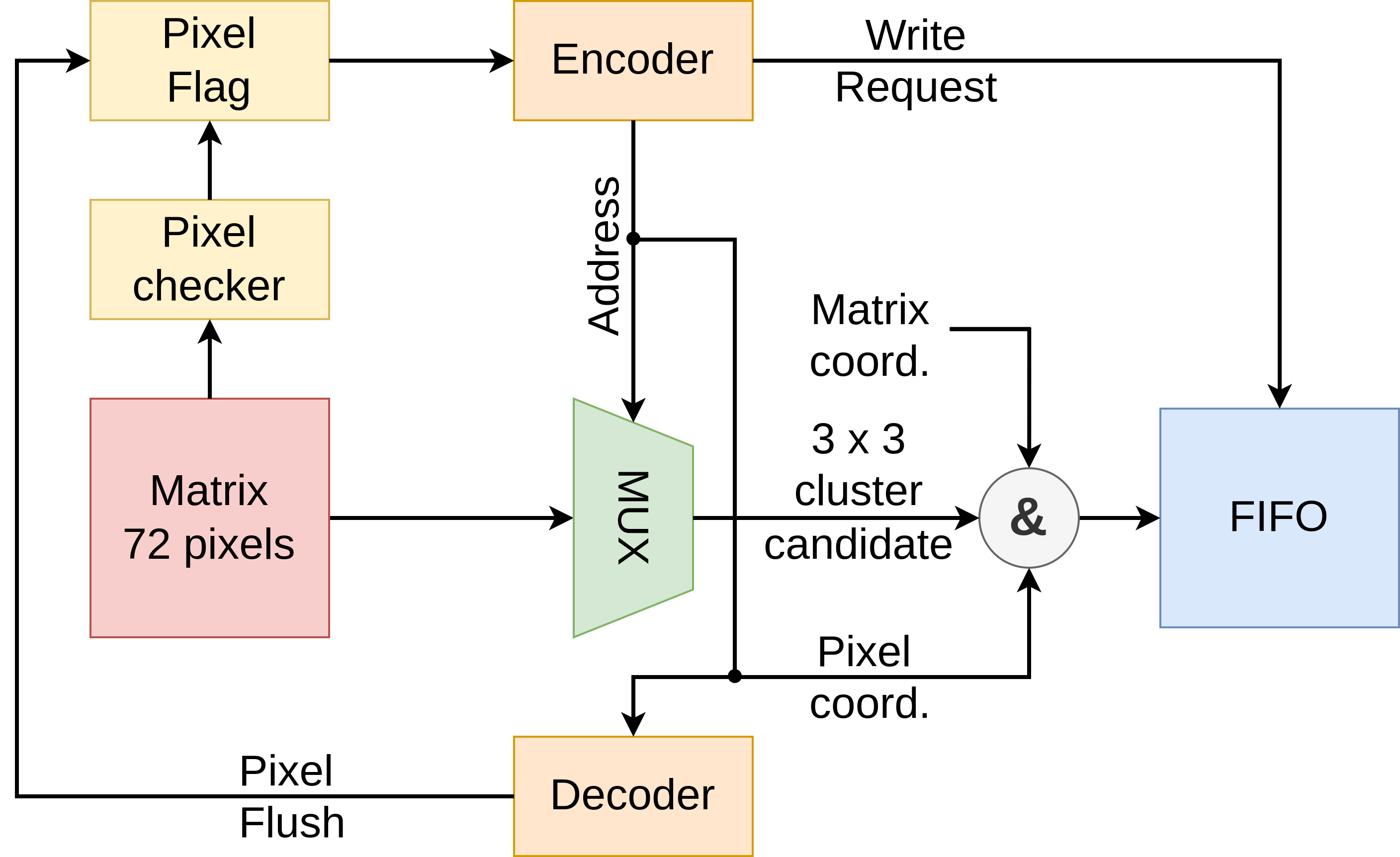}
    \caption{Cluster finder block diagram and its data flow.}
    \label{fig:cluster-finder}
\end{figure}

During the second step, the cluster finder block processes the content of the corresponding filled matrix. Figure~\ref{fig:cluster-finder} shows the logic of how clusters are reconstructed, starting from the matrix pixel content.
Each pixel in a matrix checks if it belongs to one of the L-shaped patterns of the algorithm through the pixel checker block. This process is performed in parallel at full speed for each pixel in each matrix. When a pattern match is found, an anchor pixel in the matrix is identified. As a consequence, the bit in the pixel flag vector corresponding to the position of the anchor pixel in the matrix is set to 1. An encoder reads the pixel flag vector content and passes the addresses of all found anchor pixels to a multiplexer, one at a time. The multiplexer extracts the 3$\times$3 cluster candidate corresponding to the address received from the encoder. As soon as an anchor pixel has been processed and corresponding cluster candidate found, the decoder block receives the pixel address from the encoder and resets the corresponding bit to zero in the pixel flag vector. The reset operation is performed by means of the pixel flush signal. For each cluster, a word containing the matrix coordinates, the anchor pixel position and the 3$\times$3 cluster candidate is written in the matrix FIFO. A merger reads the cluster candidates from all the matrix FIFOs and sends them to a LUT, which computes the centroid of each cluster (Fig.~\ref{fig:matrix-chain}). The cluster position is obtained by combining the matrix position in the detector, the anchor-pixel position in the matrix and the LUT output. The cluster words are then saved into a FIFO that contains all the clusters from non-isolated SPs of a VELO sensor that do not overflow the matrix chain. The two data lines at the end of the matrix chain which carry overflow SPs are merged into a single line. Overflow SPs are reconstructed as if they were isolated by means of a LUT, and the reconstructed clusters are stored into a FIFO.

\subsection{Encoder}

\begin{figure}[tb]
    \centering
    \includegraphics[width=15pc]{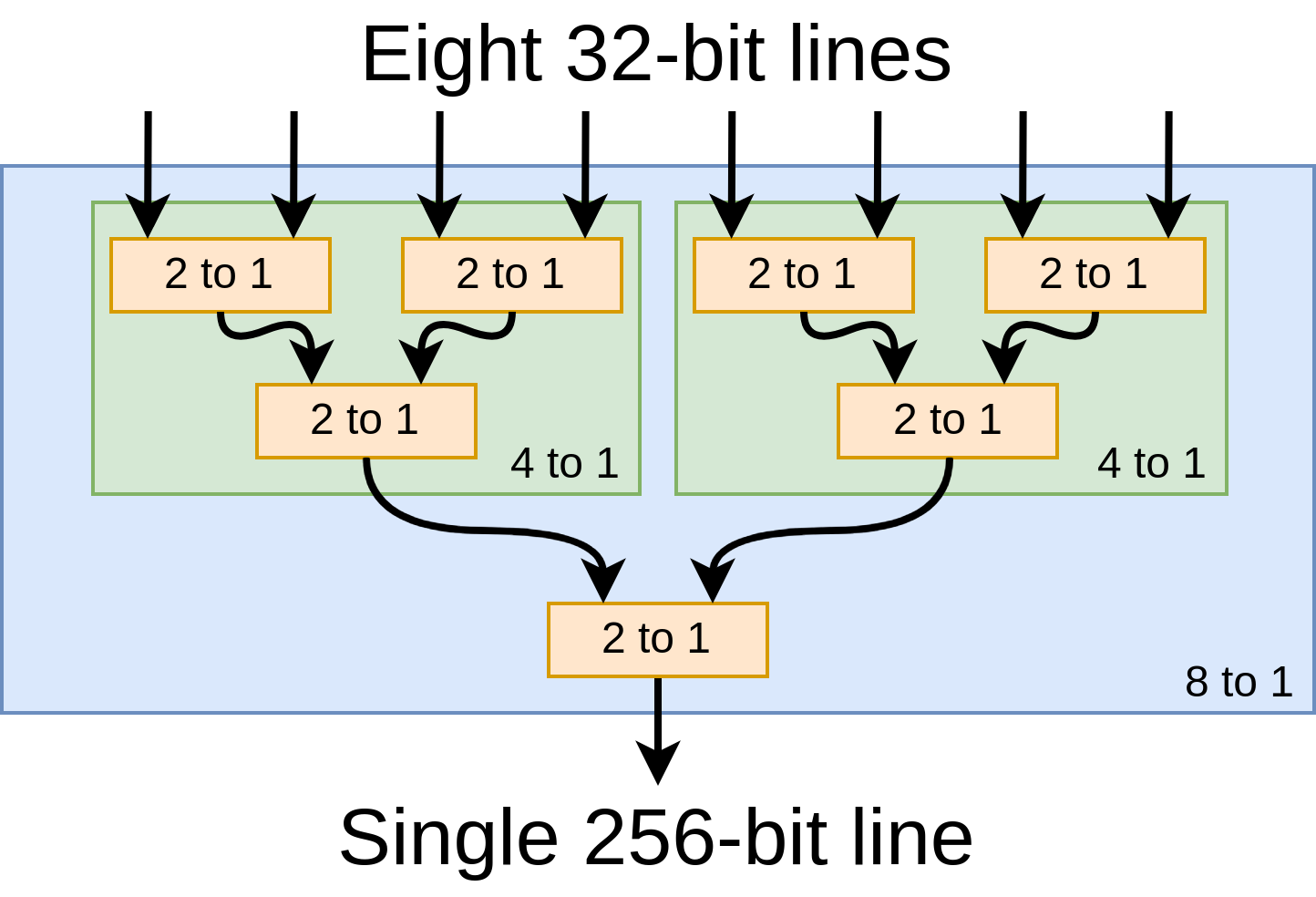}
    \caption{Structure of a 8-to-1 encoder built from 2-to-1 encoders.}
    \label{fig:encoder-recursive}
\end{figure}

\begin{figure}[tb]
    \centering
    \includegraphics[width=20pc]{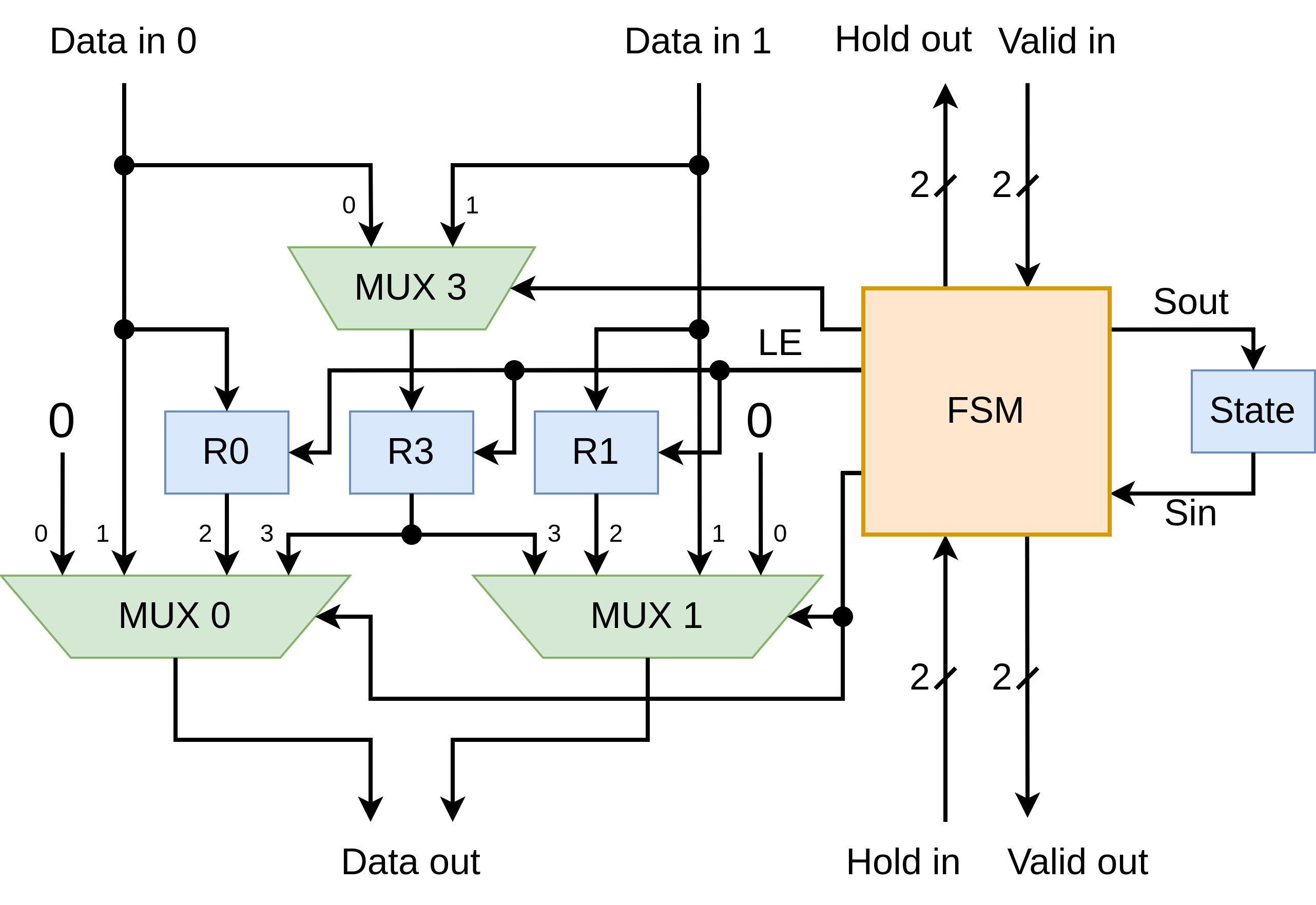}
    \caption{2-to-1 encoder block diagram. R0, R1, R3 and State are registers, MUX0, MUX1 and MUX3 are multiplexers and FSM is a finite state machine that manages hold, valid and latch enable (LE) write signals.}
    \label{fig:encoder}
\end{figure}

The last processing block of the clustering architecture is devoted to encoding the eight separate 32-bit data streams into a single 256-bit bus, to comply with the required output format. The encoder architecture has been designed as a trade-off between speed and bandwidth optimisation. The encoder is required to output a 256-bit word at each clock cycle, to maintain a throughput larger than 30\mhz. Given the speed constraint, the SP packing performed by the encoder is not optimal in each event, interleaving zero-padded words in between 256-bit words to match the output width.
To build the complete 8-to-1 encoder, seven 2-to-1 encoders are instantiated, as shown in Fig.~\ref{fig:encoder-recursive}.
The 2-to-1 encoder block puts together two input data lines (N + N bits) into a single output (2N bits) by means of buffer registers (R0, R1 and R3) and a control FSM (Fig.~\ref{fig:encoder}). If two cluster words are received and no hold signal is asserted by the subsequent block, the two words are packed together and sent out. If a single cluster is received, it is stored in the R3 register and matched with the next input cluster. If a hold signal is received the incoming cluster is stored in the R0 or R1 register depending on its input line. In case an odd number of words is received within an event, a zero-padded word is added to match the 2N output width. When two EE signals are received, they are compared and, if they match, sent out. Otherwise, an error signal is generated.

\subsection{Monitoring and error handling}

The clustering architecture has several blocks whose behaviour affects the functioning of the entire data processing chain. Therefore, a monitoring procedure is implemented to probe each block throughout the whole reconstruction process to ensure a correct data handling. Between each block of the diagram illustrated in Fig.~\ref{fig:firmware-structure} a FIFO is inserted as a buffering element. It decouples the data writing process of the previous input block from the data reading of the subsequent output block, absorbing local processing rate fluctuations.

The occupancy levels of all interposed FIFOs, as well as their maxima over a certain time interval, are periodically read to check for, and diagnose, possible slowdowns of any processing blocks. The fraction of SPs overflowing the matrix chain is also monitored. 
Each processing block is also equipped with an error-checking logic, which monitors two types of errors. The first type corresponds to a data loss, occurring when a block receives valid data in input and the register in which data should be written is already full. The second type occurs when mismatching EE signals are received, indicating a loss of synchronisation in the input data. In both cases, a signal is generated and an error word is output, containing a code to trace back the origin of the error for debugging purposes. A reset signal needs to be sent to the clustering logic and memories to recover from both error types.

\section{FPGA resource usage and throughput}
\label{sec:resource_throughput}

The clustering architecture was initially compiled and tested standalone on a Stratix~V based prototyping board \cite{dini-brd}. The FPGA device mounted on the prototyping board has similar amount of logic, memory resources and clock speed to the Arria 10 carried by the TELL40 readout boards. During the test, the firmware was fed with simulated SP data from RAM memories that are read in a loop. The clusters reconstructed in hardware were compared to the output of the high-level C++ simulation of the algorithm, run on the same set of input SPs. The quality of the reconstruction and the reliability of the measurements were verified. 

The firmware can process events with up to an average of 32 SPs per VELO half-module, using a 350\mhz clock rate. This condition is met for the whole VELO detector, where the average occupancy is 26 SPs per event, near the nominal interaction point.
An average event processing rate of 38.9\mhz is measured on minimum-bias LHC collision events, in the VELO module with the highest occupancy. The measurement is also performed on $pp$ collisions with higher than average track multiplicity, containing reconstructible $B^0_s \rightarrow \phi\phi$ decays, as a sample of typical data that the LHCb DAQ would select and save on permanent storage. The measured throughput of 30.9\mhz is still higher than the average LHC bunch crossing rate, and ensures that even a random fluctuation leading to the occurrence of several high-occupancy events in a row poses no risk of clogging the pipeline. The clustering firmware is therefore expected to run safely throughout the entire Run~3 physics data taking.

Compiling the entire VELO firmware within the Arria 10 allows for the measurement of the amount of resources needed to perform clustering in real time. The clustering firmware requires roughly 31\% of the logic and 11\% of the memory of an Arria 10 chip to process an entire VELO module.
After standalone validation, the clustering firmware was combined and fully integrated with the readout firmware to build the complete VELO readout firmware. Additional features were added in the integration process, like the handling of global LHCb control signals, response to errors, and an optional bypass that allows both SP and cluster data to be output for debugging purposes. SPs are then fed to the LHCb simulation that outputs software-based clusters which are then compared to firmware-based ones.  The optional bypass option will be enabled periodically, or in case of a need to debug, during data taking to check that the firmware is reconstructing clusters correctly.
The overall final chip occupancy turns out to be about 75\%. Some tuning was required to fix timing violations occurring due to the large fraction of resource usage and to the complex connectivity of the design. The complete firmware was then compiled and loaded on the LHCb readout boards, and successfully tested within the DAQ system by means of signal injection in the detector front-end. The firmware has also been tested on the detector readout boards using an internal front-end generator within the firmware, capable of generating input data at the nominal data rate (64\,Gb/s).  At the time of this writing, the firmware is fully commissioned and has started to take physics data in LHC Run~3.
\begin{figure}[tb]
    \centering
    \includegraphics[width=20pc]{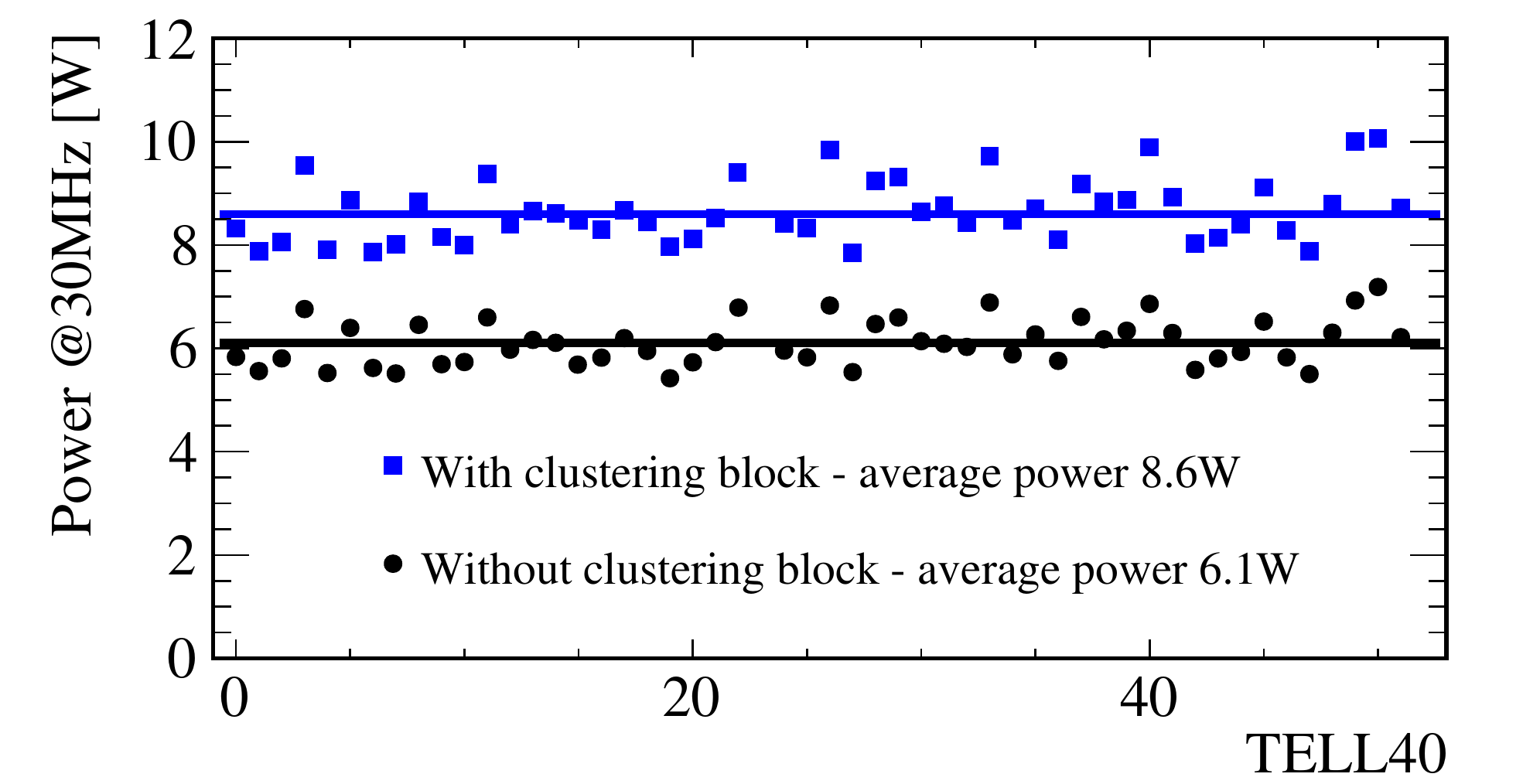}\\
    \includegraphics[width=20pc]{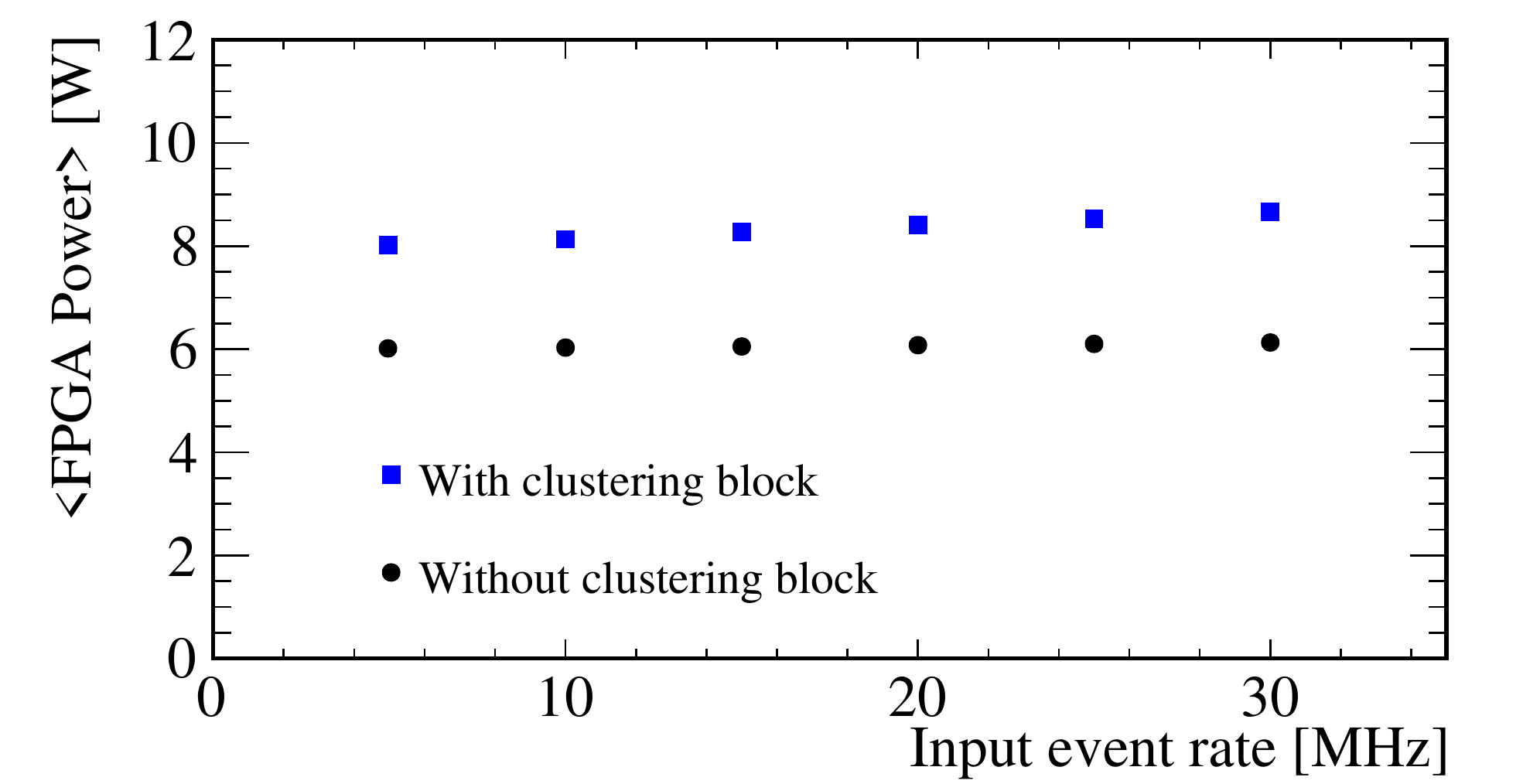}
    \caption{Power consumption of individual VELO TELL40 FPGAs (top) processing data at an event input rate of 30 MHz. The average value over the 52 FPGAs is also superimposed with an horizontal line. Average power consumption over all 52 FPGAs as a function of the input event rate (bottom). Measurements  using the firmware without the cluster finding block (outputting SPs instead of clusters) are also reported for comparison.}
    \label{fig:power}
\end{figure}

The FPGA power consumption of all VELO TELL40s is measured at the nominal event rate of 30\,MHz (Fig.~\ref{fig:power}). For comparison, the same measurements are performed using the firmware without the cluster finding block, outputting SPs instead of clusters. The average power consumption of an FPGA within a TELL40 card when processing an event rate of 30\,MHz with the readout firmware only is 6.1\,W; this increases to a total of 8.6\,W for the full firmware, including the clustering block. The same measurements are repeated for different values of the input event rate, showing a very slow increase of power consumption with the input rate.

\section{Summary and Conclusions}
\label{sec:summary}

A novel two-dimensional clustering architecture was developed, implemented in the VHDL language, and integrated in the LHCb readout FPGA cards. The architecture exploits the principles developed within the INFN-RETINA R\&D project~\cite{cenci-twepp} for real-time track reconstruction, and effectively represents its first processing stage.

This firmware proved capable of directly processing every event at the 30\mhz LHC crossing rate (a total flow of 5\,Tb/s) without time-multiplexing or buffering of any sort, in a manner that serves the needs of an actual high-energy physics experiment. The physics performances of the algorithm were extensively studied and showed to be effectively indistinguishable from software clustering algorithms. The sparse-matrix technique adopted in its implementation proved successful in handling large detectors (order of 40 million pixels) with a modest amount of logic and memory resources. This allowed its insertion into the existing LHCb readout hardware, for use in the Run~3 physics data taking.
This is a significant advancement over the previous state-of-the-art in HEP. The previous best performing cluster-finding system implemented in FPGAs has a throughput of about 100\khz and requires the deployment of about 4~parallel firmware copies, processing about 15~MPixel/s each~\cite{annovi-atlas}.

Moving the VELO clustering reconstruction from the HLT1 sequence to the FPGA readout cards leads to a measurable throughput improvement.
Without accounting for isolation flagging, for which no software implementation is available for comparison, the present cluster finder firmware allows a savings of about 11\% of the computing power of the LHCb HLT1 full reconstruction sequence, allowing a corresponding increase of the LHCb data-taking rate. As a further advantage, a reduction of the VELO data size of approximately~14\% was obtained, which allows to save resources both in the DAQ chain and in permanent data storage. Details on the GPU-based VELO clustering reconstruction can be found in Ref.~\cite{campora-thesis}.
In addition, the FPGA implementation consumes significantly less electrical power than its GPU analogue. From the data in Fig.~\ref{fig:power}, it follows that the set of 52 VELO TELL40s requires about 130~W of power to perform cluster reconstruction of the entire VELO, while the GPU implementation would require about 6~kW (again not including isolation flagging). The power needed to perform cluster reconstruction on GPUs is estimated by multiplying the GPU power usage (230~W) by the number of GPUs (236) required to process a 30~MHz input event rate and by the fraction of time spent in cluster reconstruction (11\%). This is also in agreement with the measurements presented in Ref.~\cite{RTA-power}.

\section{Further considerations}
\label{sec:CCL}

In a broader perspective, this work can be seen as a special case of Connected Component Labelling with Centre of Gravity calculation (COG) -- a computation that often occurs in image processing systems with the purpose of identifying connected sets of pixels belonging to the same visual feature. 
The main difference is the modest size of the features of our interest, that we could contain within a 3$\times$3 matrix, and their sparseness, that makes our problem somewhat simpler. However, this greater simplicity comes with a `frame rate' requirement (30\,MHz) that is orders of magnitude larger than typical image processing rates ($<$1\,kHz).  In fact, a CPU implementation exists of the same VELO clustering task discussed in this paper that was inspired by some algorithms in use in image processing problems, appropriately revisited to exploit the smallness of the size of the components and their sparseness~\cite{sparseCCL}.  

In recent years, also this type of image processing tasks is increasingly being moved from CPUs to dedicated FPGA firmware to achieve greater speed and efficiency, and it may be interesting to compare those solutions to the present work. 
As an illustrative example, we take the FPGA implementation described in Ref.~\cite{ccl-embedded}. There, frames of 640$\times$480 pixels are processed at a rate of 730\,Hz, by a Zynq AP-SOC 7045 FPGA, running a 225\,MHz clock, without COG.  This system compares well with our case, where each of the 104 instances of our firmware processes a matrix of 512$\times$768 pixels, and our clock frequency and resource usage are also quite similar. The Arria 10 FPGA mounted on TELL40 cards has a capacity of 1150k logic elements whereas the Zynq 7045 FPGA has 350k logic cells. A single instance of the clustering firmware requires about 15\% of the available logic on the chip, while for the studies reported in Ref.~\cite{ccl-embedded} we assume a typical usage of about 50\% of the total resources.
However, our frame rate is larger by a huge factor, of nearly $10^5$. This difference is likely due to the sequential structure of the image processing firmwares, that proceeds by a raster scan rather than by a massively parallel calculation; but is definitely also a consequence of the greater simplicity of our problem in terms of cluster size and occupancy. 
In fact, cases of FPGA-based CCL implementations that reach a throughput comparable to that of our architecture are based on breaking down the image in smaller parts that are analysed in parallel, and later coalesced\cite{CCL-fpga}; an approach that bears some resemblance to our use of sparse matrices. 

However, all the above examples assume that the image data arrive as an ordered sequence of pixels, and do not provide detailed topology analysis of the found clusters, so they could not be straightforwardly applied to our problem.  Conversely, the smallness of the components addressed by our system may not be of interest in general image processing applications; nevertheless, it cannot be excluded that some of the ideas described in this article could find some use in image processing tasks, at least in some specific instances.

\section*{Acknowledgement}
We are thankful for the support of the LHCb Real-Time Analysis group, within which this project was developed. A special thank goes to the LHCb VELO group for the tight collaboration, support, and integration coordination, without which this work would not have been possible. The authors would also like to thank the LHCb computing and simulation teams for providing the simulated samples used in the paper.

\bibliographystyle{IEEEtran}
\bibliography{bibliography}

\begin{thebibliography}{10}
\providecommand{\url}[1]{#1}
\csname url@samestyle\endcsname
\providecommand{\newblock}{\relax}
\providecommand{\bibinfo}[2]{#2}
\providecommand{\BIBentrySTDinterwordspacing}{\spaceskip=0pt\relax}
\providecommand{\BIBentryALTinterwordstretchfactor}{4}
\providecommand{\BIBentryALTinterwordspacing}{\spaceskip=\fontdimen2\font plus
\BIBentryALTinterwordstretchfactor\fontdimen3\font minus
  \fontdimen4\font\relax}
\providecommand{\BIBforeignlanguage}[2]{{%
\expandafter\ifx\csname l@#1\endcsname\relax
\typeout{** WARNING: IEEEtran.bst: No hyphenation pattern has been}%
\typeout{** loaded for the language `#1'. Using the pattern for}%
\typeout{** the default language instead.}%
\else
\language=\csname l@#1\endcsname
\fi
#2}}
\providecommand{\BIBdecl}{\relax}
\BIBdecl

\bibitem{online-upgrade}
\BIBentryALTinterwordspacing
{\relax LHCb Collaboration}, ``{LHCb Trigger and Online Upgrade Technical
  Design Report},'' CERN, Geneva, Tech. Rep. CERN-LHCC-2014-016, 2014.
  [Online]. Available: \url{https://cds.cern.ch/record/1701361}
\BIBentrySTDinterwordspacing

\bibitem{upgrade2-eoi}
\BIBentryALTinterwordspacing
{\relax LHCb Collaboration}, ``{Expression of Interest for a Phase-II LHCb
  Upgrade: Opportunities in flavour physics, and beyond, in the HL-LHC era},''
  CERN, Geneva, Tech. Rep., Feb 2017. [Online]. Available:
  \url{https://cds.cern.ch/record/2244311}
\BIBentrySTDinterwordspacing

\bibitem{GPU-HLT}
\BIBentryALTinterwordspacing
{\relax LHCb Collaboration}, ``{LHCb Upgrade GPU High Level Trigger Technical
  Design Report},'' CERN, Geneva, Tech. Rep. CERN-LHCC-2020-006, May 2020.
  [Online]. Available: \url{https://cds.cern.ch/record/2717938}
\BIBentrySTDinterwordspacing

\bibitem{cenci-twepp}
\BIBentryALTinterwordspacing
R.~Cenci \emph{et~al.}, ``{\relax Development of a High-Throughput Tracking
  Processor on FPGA Boards},'' in \emph{Proc. Topical Workshop on Electronics
  for Particle Physics (TWEPP 2017)}, Santa Cruz, CA, USA, 2017. [Online].
  Available: \url{https://pos.sissa.it/313/136/}
\BIBentrySTDinterwordspacing

\bibitem{velo-upgrade}
\BIBentryALTinterwordspacing
{\relax LHCb Collaboration}, ``{LHCb VELO Upgrade Technical Design Report},''
  CERN, Geneva, Tech. Rep. CERN-LHCC-2013-021. LHCB-TDR-013, 2013. [Online].
  Available: \url{https://cds.cern.ch/record/1624070}
\BIBentrySTDinterwordspacing

\bibitem{firmware-repo}
\BIBentryALTinterwordspacing
G.~Bassi \emph{et~al.}, ``{FPGA implemention of a fast 2D clustering algorithm
  (VHDL language)},'' 2019. [Online]. Available:
  \url{https://doi.org/10.15161/oar.it/23524}
\BIBentrySTDinterwordspacing

\bibitem{velopix}
\BIBentryALTinterwordspacing
T.~Poikela \emph{et~al.}, ``{\relax VeloPix: the pixel ASIC for the LHCb
  upgrade},'' \emph{JINST}, vol.~10, no.~01, p. C01057, 2015. [Online].
  Available: \url{https://doi.org/10.1088/1748-0221/10/01/C01057}
\BIBentrySTDinterwordspacing

\bibitem{velo-firmware}
\BIBentryALTinterwordspacing
K.~Hennessy \emph{et~al.}, ``{\relax Readout Firmware of the Vertex Locator for
  LHCb Run 3 and Beyond},'' \emph{IEEE Transactions on Nuclear Science},
  vol.~68, no.~10, pp. 2472--2479, 2021. [Online]. Available:
  \url{https://cds.cern.ch/record/2789034}
\BIBentrySTDinterwordspacing

\bibitem{tell40}
\BIBentryALTinterwordspacing
J.~P. Cachemiche \emph{et~al.}, ``{The PCIe-based readout system for the LHCb
  experiment},'' \emph{JINST}, vol.~11, p. P02013. 12 p, 2016. [Online].
  Available: \url{http://cds.cern.ch/record/2262859}
\BIBentrySTDinterwordspacing

\bibitem{simulation}
\BIBentryALTinterwordspacing
S.~Miglioranzi \emph{et~al.}, ``{The LHCb Simulation Application, Gauss:
  Design, Evolution and Experience},'' CERN, Geneva, Tech. Rep.
  LHCb-PROC-2011-006. CERN-LHCb-PROC-2011-006, Jan 2011. [Online]. Available:
  \url{https://cds.cern.ch/record/1322402}
\BIBentrySTDinterwordspacing

\bibitem{luca-thesis}
\BIBentryALTinterwordspacing
L.~Giambastiani, ``{A 2D FPGA-based clustering algorithm for the LHCb silicon
  pixel detector running at 30 MHz},'' Master's thesis, {Università di Pisa,
  Pisa, IT}, 2020, presented 16 Jul 2020. [Online]. Available:
  \url{https://cds.cern.ch/record/2725831}
\BIBentrySTDinterwordspacing

\bibitem{giovanni-thesis}
\BIBentryALTinterwordspacing
G.~Bassi, ``{A FPGA-based architecture for real-time cluster finding in the
  LHCb silicon pixel detector},'' Ph.D. dissertation, {Scuola Normale
  Superiore, Pisa, IT}, 2023. [Online]. Available:
  \url{https://cds.cern.ch/record/2845901}
\BIBentrySTDinterwordspacing

\bibitem{heavy-ion}
\BIBentryALTinterwordspacing
R.~Litvinov, ``{\relax LHCb: Heavy-ion physics results and prospects},''
  \emph{International Journal of Modern Physics E}, vol.~30, no.~11, p.
  2141004, 2021. [Online]. Available: \url{https://cds.cern.ch/record/2804032}
\BIBentrySTDinterwordspacing

\bibitem{tracker-upgrade}
\BIBentryALTinterwordspacing
{\relax LHCb Collaboration}, ``{LHCb Tracker Upgrade Technical Design
  Report},'' CERN, Geneva, Tech. Rep. CERN-LHCC-2014-001, 2014. [Online].
  Available: \url{http://cds.cern.ch/record/1647400/}
\BIBentrySTDinterwordspacing

\bibitem{tracking-defs}
\BIBentryALTinterwordspacing
{\relax LHCb Collaboration}, ``{Tracking Definitions and Conventions for Run 3
  and Beyond},'' CERN, Geneva, Tech. Rep., Feb 2021. [Online]. Available:
  \url{https://cds.cern.ch/record/2752971}
\BIBentrySTDinterwordspacing

\bibitem{dini-brd}
{\relax Dini Group\textsuperscript{\textregistered}},
  \url{https://www.synopsys.com/verification/prototyping/dini-products.html},
  {\relax Board model: DNS5GX\_F2}.

\bibitem{annovi-atlas}
\BIBentryALTinterwordspacing
C.-L. Sotiropoulou \emph{et~al.}, ``{\relax A Multi-Core FPGA-Based
  2D-Clustering Implementation for Real-Time Image Processing},'' \emph{IEEE
  Transactions on Nuclear Science}, vol.~6, 12 2014. [Online]. Available:
  \url{https://doi.org/10.1109/TNS.2014.2364183}
\BIBentrySTDinterwordspacing

\bibitem{campora-thesis}
\BIBentryALTinterwordspacing
D.~H. C{\'a}mpora~P{\'e}rez, ``Optimization of high-throughput real-time
  processes in physics reconstruction,'' Ph.D. dissertation, Universidad de
  Sevilla, 2019, \relax VELO clustering is discussed in chapter 3.1. [Online].
  Available: \url{http://cds.cern.ch/record/2718278}
\BIBentrySTDinterwordspacing

\bibitem{RTA-power}
\BIBentryALTinterwordspacing
R.~Aaij \emph{et~al.}, ``{Evolution of the energy efficiency of LHCb’s
  real-time processing},'' \emph{EPJ Web Conf.}, vol. 251, p. 04009, 2021.
  [Online]. Available: \url{https://cds.cern.ch/record/2773126}
\BIBentrySTDinterwordspacing

\bibitem{sparseCCL}
\BIBentryALTinterwordspacing
A.~Hennequin \emph{et~al.}, ``{SparseCCL: Connected Components Labeling and
  Analysis for sparse images},'' in \emph{{DASIP 2019 - The Conference on
  Design and Architectures for Signal and Image Processing}}, Montr{\'e}al,
  Canada, 2019. [Online]. Available:
  \url{https://hal.archives-ouvertes.fr/hal-02343598}
\BIBentrySTDinterwordspacing

\bibitem{ccl-embedded}
\BIBentryALTinterwordspacing
F.~Spagnolo \emph{et~al.}, ``{\relax An Efficient Connected Component Labeling
  Architecture for Embedded Systems},'' \emph{Journal of Low Power Electronics
  and Applications}, vol.~8, no.~1, 2018. [Online]. Available:
  \url{https://www.mdpi.com/2079-9268/8/1/7}
\BIBentrySTDinterwordspacing

\bibitem{CCL-fpga}
M.~J. Klaiber \emph{et~al.}, ``{\relax A high-throughput FPGA architecture for
  parallel connected components analysis based on label reuse},'' in \emph{2013
  International Conference on Field-Programmable Technology (FPT)}, Kyoto,
  Japan, 2013.

\end{thebibliography}

\end{document}